\pdfoutput=1

\documentclass[12pt]{article}
\usepackage{jheppub}

\usepackage{caption}
\usepackage{subcaption}

\usepackage[latin9]{inputenc}
\usepackage{float}
\usepackage{esint}

\usepackage{amsmath}
\usepackage{latexsym,amssymb}
\usepackage{graphicx,color,slashed}
\usepackage{bbm} 
\usepackage{mathtools}
\usepackage{ifpdf}

\newcommand{\cF}{\mathcal{F}}
\newcommand{\cG}{\mathcal{G}}

\newcommand{\cC}{\mathcal{C}}

\newcommand{\cH}{\mathcal{H}}
\newcommand{\cL}{\mathcal{L}}
\newcommand{\cM}{\mathcal{M}}
\newcommand{\cN}{\mathcal{N}}
\newcommand{\cO}{\mathcal{O}}

\newcommand{\cV}{\mathcal{V}}

\newcommand{\be}{\begin{equation}}
\newcommand{\ee}{\end{equation}}
\newcommand{\ba}{\begin{eqnarray}}
\newcommand{\ea}{\end{eqnarray}}

\renewcommand{\a}{\alpha}
\renewcommand{\l}{\lambda}
\renewcommand{\b}{\beta}
\newcommand{\N}{\mathcal{N}}

\def\E{{$E_{7(7)}$}}

\newcommand{\rf}[1]{(\ref{#1})}
\newcommand{\bea}{\begin{eqnarray}}
\newcommand{\eea}{\end{eqnarray}}

\def\bfzero{\relax{\rm I\kern-.18em 0}}
\def\bfone{\relax{\rm 1\kern-.35em 1}}
\def\twomat#1#2#3#4{\left(\begin{array}{cc}
\end{array}
\right)}

\def\cC{{\cal C}}

\usepackage[mathscr]{euscript}

\title{\rm{\bf  {Deformation of   $ d=4, \, \, \N\geq 5$  Supergravities Breaks  Nonlinear Local Supersymmetry}}}
\author[a]{ Renata Kallosh}
\author[b]{ and Yusuke Yamada}
\affiliation[a]{Stanford Institute for Theoretical Physics and Department of Physics, Stanford University, Stanford, CA 94305, USA}
\affiliation[b]{Waseda Institute for Advanced Study, Waseda University, 1-21-1 Nishi Waseda, Shinjuku, Tokyo 169-0051, Japan}

\emailAdd{kallosh@stanford.edu}
\emailAdd{y-yamada@aoni.waseda.jp}
\abstract{ We study   $ d=4,  \, \N\geq 5$  supergravities and  their deformation via candidate counterterms, with the purpose  to absorb UV divergences. We generalize the earlier studies of deformation and twisted self-duality constraint to the case with unbroken local $\cH$-symmetry in presence of fermions.  We find that the deformed action breaks nonlinear local supersymmetry. We show that all known cases of enhanced UV divergence cancellations are explained by nonlinear local supersymmetry.

This result implies, in particular,  that if $\N=5$ supergravity at five loop will turn out to be UV divergent, the deformed theory will be BRST inconsistent. If it will be 
finite, it will be a consequence of nonlinear local supersymmetry and E7-type duality. 
  }

\begin{document}

\maketitle



\parskip 5pt


\section{Introduction} 
\subsection{Deformation of  theories with local symmetries and BRST symmetry}
 There are two main issues one may address in perturbative quantum theories of gravitational fields: 

1. What kind of UV divergences are predicted  for the loop computations using original undeformed classical theory on the basis of symmetries of this theory? 

2. Is it possible to deform the theory {\it consistently} by adding the higher derivative terms which absorb UV divergences and introduce new couplings beyond  the gravitational coupling $\kappa^2= {1\over M_{Pl}^2}$?  

``Consistently'' here has a meaning that local symmetries might be deformed but the action has to be  invariant under deformed local symmetries. Also in supergravity the global E7-type duality symmetries might be deformed, but the deformed action needs to be duality invariant on shell, to support  unitarity.  

In pure gravity  the answer to both of these problems  is known.
A  2-loop UV  divergence  $R^3$  was predicted in  \cite{Kallosh:1974yh} and confirmed by computations in 
\cite{Goroff:1985th}. The 1-loop UV divergence in the form of a topological Gauss-Bonnet  term was revealed by the computation in 
\cite{Gibbons:1978ac}.  At present, a pure gravity action, which provides finite amplitudes  up to 2-loop order can be  viewed as a deformed Einstein-Hilbert action. For example,  in \cite{Abreu:2020lyk} we find
\be
{\cal L}_{\rm deformed}^{\rm gravity}=  -{2\over \kappa^2} \sqrt{|g|} R + {\cC_{\rm GB}\over (4\pi)^2} \sqrt{|g|} R_{ \mu\nu \rho\sigma}^* R^{*\mu\nu \rho\sigma} + {\cC_{R^3}\over (4\pi)^2}\Big ({\kappa\over 2}\Big )^2 \sqrt{|g|} R_{\alpha \beta}{}^{\mu\nu} R_{\mu \nu}{}^{\rho\sigma} R_{\rho\sigma}{}^{\alpha \beta}+\dots \ .
\label{deformed}\ee
Here the new couplings absorb the known UV divergences
\be
\cC_{GB}= \Big ( {53\over 90 \epsilon} + c_{\rm GB} (\mu) \Big ) \mu^{-2\epsilon} \ ,
\qquad 
\cC_{R^3}= \Big ( {209\over 1440 \epsilon} + c_{R^3} (\mu) \Big ) \mu^{-4\epsilon} .
\label{coupl}\ee
The renormalized couplings $c_{\rm GB} (\mu)$  and $c_{R^3} (\mu)$ are new parameters describing pure gravity.  Each of the 3 parts of the action \rf{deformed} is separately invariant under the  off-shell gauge symmetry 
\be
\delta g_{\mu\nu}= \nabla_\mu \xi_\nu + \nabla_\nu \xi_\mu \ .
\label{gs}\ee
The gauge symmetry of the classical theory is not deformed by quantum corrections requiring the deformation of the Einstein-Hilbert  action.

The prediction of the $R_{\alpha \beta}{}^{\mu\nu} R_{\mu \nu}{}^{\rho\sigma} R_{\rho\sigma}{}^{\alpha \beta}$ 2-loop UV divergence in \cite{Kallosh:1974yh} was based on a covariant formalism where the simplest form of Ward Identities is valid.
In such case the UV divergence can be predicted to form an invariant under the gauge symmetry of the classical action. An analogous conclusion follows from BRST symmetry defining 
perturbative QFT in a  consistent  gauge theory  \cite{Becchi:1975nq,Tyutin:1975qk}. The first step in BRST construction is the existence of the  local action, classical or deformed,  invariant under local symmetry, classical or deformed. 
Namely,  iff the  deformed by a counterterm (CT)  action
\be
S^{\rm def} = S_{\rm cl} + \l S^{\rm CT} 
\label{Def} \ee
 has a local (classical or deformed) gauge symmetry, we can call it $ S_{\rm inv}$. In such case
 one can add to this action some gauge fixing terms, as  well as the required ghosts action, \be
 S^{\rm BRST}  = S_{\rm inv} + S_{\text{gauge-fixing}} + S_{\rm ghosts}\ee
 and prove the  symmetry under BRST transformation $Q^{\rm BRST}$ of the total action $S^{\rm BRST}$  which controls quantum corrections
 \be
Q^{\rm BRST} \,  S^{\rm BRST}=0\, , \qquad Q_{\rm BRST}^2=0.
 \ee
When the deformation terms break some of the local symmetries of the classical action, so that $S^{\rm def}$ is not $ S_{\rm inv}$, 
 the BRST construction based on $S^{\rm def}$ becomes  inconsistent. The proof of gauge symmetry and unitarity of the perturbative gauge theory becomes invalid, like in theories with gauge anomalies.

In gravity,  the action  \rf{deformed} is invariant under  the  gauge symmetry \rf{gs}, one can construct a BRST action so that quantum corrections  control  the loop computations in a consistent perturbative QFT.

In $d=4, \cN=8$ supergravity  \cite{Cremmer:1979up,deWit:1982bul},  as well as in all pure (no matter)  $\cN$-extended supergravities,
the possible UV divergences were predicted in the past \cite{Kallosh:1980fi,Howe:1980th} on the basis of a Lorentz-covariant on shell superspace geometry \cite{Brink:1979nt, Howe:1981gz}. $\cN=8$ supergravity was also studied in the light-cone off shell superspace \cite{Brink:2008qc} where \E\ symmetry  commutes with the super-Poincar\'e group.

The issue of UV divergences was consequently revisited in  $\cN=8$ 
 light-cone superspace in  \cite{Kallosh:2009db,Kallosh:2010kk}. It was found that all  candidate CT' s proposed in \cite{Kallosh:1980fi,Howe:1980th} are ruled out since they are not available in the off-shell  light-cone superspace.
 
 However, at smaller $\cN=5,6$, where loop computations are also possible, the arguments in \cite{Kallosh:2009db,Kallosh:2010kk} are difficult to apply. Here and hereafter, we will focus only on $d=4$ unless otherwise noted. The light-cone superspace is complicated even in maximal supergravity \cite{Brink:2008qc},  and in $\cN=5,6$ it was not developed. The Lorentz-covariant superspace \cite{Howe:1981gz}, as well as supergravity actions at $\cN=5,6$, are better known from the consistent truncation of maximal supergravity. Therefore to study all $\cN=5,6,8$ supergravities and their UV divergences we proceed with point 2 above, where candidate CTs are known from the on-shell Lorentz-covariant superspace \cite{Kallosh:1980fi,Howe:1980th}.
 
The analysis of UV divergences in $\cN=5,6,8$  was already performed in \cite{Kallosh:2012yy,Gunaydin:2013pma,Kallosh:2018wzz,Gunaydin:2018kdz} using manifest E7-type symmetry or
 properties of the unitary conformal supermultiplets. Under  assumptions that there are no supersymmetry anomalies, it was predicted that these theories will be UV finite. Here we will study the effect of UV divergences on nonlinear local supersymmetry directly.

We  will  ask a question: can we expect a deformation of $\cN=5,6,8$ supergravities of the kind we see in pure gravity? Once, at some loop order, UV divergence is detected, we add the relevant expression to the original action and deform it: this term will absorb UV divergence and provide additional couplings with higher derivatives, as in eqs. \rf{deformed}, \rf{coupl} in pure gravity.

 The   goal of this paper to establish the symmetries of the deformed  $\cN=5,6,8$ supergravities, local supersymmetry, local $\cH$-symmetry and E7 duality,  and to check the consistency of such a deformation. 

 In the past, the role of a local nonlinear supersymmetry and of a local $\cH$ symmetry in $d=4$ supergravity was not emphasized enough, although both are  known to be  features of geometric CT's existing at $L\geq L_{cr} =\cN$, and both are broken at non-geometric linearized CT's at $L < L_{\rm cr} =\cN$ \cite{Kallosh:1980fi,Howe:1980th}. 

The advantage of using local $\cH$-symmetry in $d=4$ supergravity is that E7 symmetry is independent on local $\cH$-symmetry. Meanwhile in the unitary gauge, the manifest rigid $\cH$-symmetry involves an additional compensating $\cH$-symmetry transformation preserving the unitary gauge,  it is a mix of E7 with $\cH$, see  details in \cite{Kallosh:2008ic}.

In a recent review paper  \cite{Bern:2023zkg} the list of three cases of enhanced cancellation of a UV divergence at
 1) $\cN=5, L=4, d=4$, \, 2) $\cN=4, L=3, d=4$, \, 3) $\cN=4, L=2, d=5$ was given. We will show here that all these cases are explained by nonlinear local supersymmetry. The first case is just below in Sec. \ref{sec:12}, the case  of $\cN=4, L=3, d=4$ supergravity in Appendix \ref{appA}, since the main part of this  paper is about $\cN\geq 5, d=4$ and 
and the one in half-maximal 
 supergravity in $d=5$ is in a separate work \cite{RK}.

\subsection{Enhanced cancellation of UV divergence in $d=4, \cN=5,  L=4$}\label{sec:12}
The reason to discuss this case in the Introduction is the fact that during almost a decade  its only explanation was given in \cite{Kallosh:2018wzz,Gunaydin:2018kdz}. But this explanation was not specific for $d=4, \cN=5,  L=4$,  it was a prediction of UV finiteness at all loops based on duality symmetry, assuming unbroken supersymmetry. In this paper we focus on predictions from nonlinear local supersymmetry. In this spirit we provide here a simple explanation of cancellation of UV divergences in $d=4,\, \cN=5,  L=4$. It does not extend to $L>4$ directly, higher loops need an additional study. 

 In the UV finite case of $\cN=5, L=4$  \cite{Bern:2014sna} the relevant harmonic CT  
was {\it claimed to be nonlinearly supersymmetric} \cite{Bossard:2011tq}.  We will explain here why only a linearized version of it can be justified, and that the nonlinear CT is not available.

The proof of consistency of the harmonic superspace $(\cN, p, q)$ in \cite{Hartwell:1994rp} was given for Yang-Mills theory and for $\cN=1,2,3,4$ conformal supergravity. Conformal constraints of $\cN\geq 5$ Poincar\'e supergravity in the harmonic superspace were established in \cite{Hartwell:1994rp}. It was suggested there that ``in the case of Poincar\'e supergravity one needs to find the geometrical formulation of the additional constraints''. The purpose of these additional constraints is to break the super Weyl invariance down to a super Poincar\'e invariance.

However, these additional constraints were not found during the last 3 decades since this suggestion was made.  And since $\cN=5$ Poincar\'e supergravity breaks conformal symmetry at the nonlinear level\footnote{See for example \cite{Gunaydin:2013pma,Gunaydin:2018kdz} where it is explained that  linearized  supergravity is based on  representations of $SU(2,2|\cN)$ superconformal algebra.  However  nonlinear interactions of $\cN\geq 5$  supergravity break  conformal 
supersymmetry algebra $SU(2,2|\cN)$  down to $\cN\geq 5$ Poincar\'e superalgebra. These require the additional constraints in the harmonic superspace which were discussed in  \cite{Hartwell:1994rp} but not delivered since. Even in case of $\cN=4$ where superconformal theory is available, see for example \cite{Ferrara:2012ui} and references therein, the relevant harmonic superspace constraint breaking the superconformal theory to $\cN=4$ super-Poincar\'e is not available.
These additional constraints in a superspace without harmonic variables are known,  they were presented  in details in \cite{Howe:1981gz}.}
 the consistency of the nonlinear harmonic superspace of
 $\cN=5$ Poincar\'e supergravity remains unproven. The linearized CT is
\be
\kappa^{6}\int d^4x  \, (d^{16} \theta)_{ 1}{}^5 \Big (\bar \chi^{1rs}_{\dot \beta} \chi_{\alpha \, 5 rs}  \Big )^2 \sim  \kappa^{6}\int d^4x  \, d^{20} \theta \Big (W_{ijkl} \bar W^{ijkl}  \Big )^2
\, , \qquad r,s =2,3,4
\label{N5}\ee
The CT is linearly supersymmetric since the subspace of the superspace is available  at the linear level, also
the superfield $ W_{ijkl} $ of dimension zero exists only  at the linear level. At the nonlinear level both forms of this CT are non-geometric, in agreement with \cite{Kallosh:1980fi,Howe:1980th}.    This means that they break nonlinear supersymmetry.

Thus, {\it there is no CT generalizing the one in \rf{N5} to the  nonlinear version  with unbroken local  nonlinear supersymmetry} and local $U(5)$ $\cH$-symmetry. This explains  the finiteness of $\cN=5, L=4$  \cite{Bern:2014sna}.
Comparative to arguments in \cite{Kallosh:2018wzz,Gunaydin:2018kdz}, the argument above is  simple (although  it does not cover the cases with $L\geq \cN$, which are covered in \cite{Kallosh:2018wzz,Gunaydin:2018kdz} and will be studied later in this paper).

We stress here that simple explanation of the cancellation of 82 diagrams observed in  \cite{Bern:2014sna} is that the relevant CT in \rf{N5} {\it breaks nonlinear local supersymmetry, although is preserves linear supersymmetry}.

From the point of view of amplitudes, the cancellation of these 82 diagrams
 is  surprising, it was given a name  {\it enhanced ultraviolet cancellations} in  \cite{Bern:2014sna}.  In  amplitudes  there is a manifest linear supersymmetry which controls the computations.  But {\it nonlinear supersymmetry actually controlls the computations  behind the scenes} and leads to cancellation of a UV divergence at $\cN=5, L=4$.

  It remains to be seen what  happens in computations in $\cN=5, L=5$. We will study here the theoretical predictions based on nonlinear local supersymmetry.

\subsection{A short summary and assumptions of this work}\label{sec:13}

We would like to clarify our statement as a short summary of this paper ahead of detailed discussions. It will be also a set of {\it important facts/assumptions} we are using here to derive our main result.

1. We assume that $\N\geq 5, \, d=4$  supergravities have a classical action which can be deformed by a CT,  as  we show in pure gravity in \rf{deformed} where in addition to Einstein-Hilbert term we also have $R^3$ term which allows to eliminate the UV divergence of the second loop. We assume that  
 the total action is  {\it Lorentz invariant}. 
 
 2. We use the fact that (for example, in $\cN=8$) the classical action \cite{Cremmer:1979up,deWit:1982bul}  {\it  has off shell local symmetries: Lorentz symmetry, a nonlinear local supersymmetry and  local $SU(8)$-symmetry, and on shell global \E\  symmetry}.  Before local $SU(8)$-symmetry is gauge-fixed, \E\ and local $SU(8)$-symmetry are linearly realized and independent. After local $SU(8)$-symmetry is gauge-fixed, in the unitary gauge, there is a remaining rigid  $SU(8)$-symmetry, a diagonal subgroup of \E\,$\times  SU(8)$.

 3. There is a significant {\it difference between the linear supersymmetry in superamplitudes/supergravity and local nonlinear supersymmetry in supergravity}.\footnote{ We are grateful to R. Roiban for a suggestion  to clarify this issue with an understanding that  amplitude computations manifestly preserve linearized supersymmetry. The reason for enhanced cancellation in this context is that  linearized CT's   in $d=4$ at $L\geq \cN$  can be promoted to nonlinear level, whereas the ones  at $L< \cN$ cannot. $\cN=5, L=4$ is an  example of a linear CT which has no nonlinear generalization,  which is the reason for the mysterious cancellation  of the sum of 82 diagrams in  \cite{Bern:2014sna}.} 
 In the linear supersymmetry there are certain constraints which permit the existence of the subspaces of the superspace and superfields depending only on Grassmann coordinates of the subspace.
  However, in the nonlinear case the integrability condition for  these constraints is not valid, as one can see via a  local supersymmetry algebra \cite{Brink:1979nt, Howe:1981gz}. 
 
 For example, a linearized superfield chirality condition $D_{\dot \a \, i} X=0$ has an integrability requirement breaking chirality condition, in general, for  $\cN\geq3$ where spin 1/2 fields are present in geometry and induce the torsion
 \be
\{ D_{\dot \a \, i}, D_{\dot \b \, j} \}  \, X= T_{ \dot \a \dot \b \, ijk }^{ \gamma} \, D^{k}_{ \gamma} \, X+\dots  \qquad T_{\dot \a \dot \b \, ijk}^{ \gamma}= \epsilon_{\dot \a \dot \b} \,  \chi_{ijk}^{ \gamma} .
\label{torsion} \ee
 It follows that at the nonlinear level a chiral superfield must be a constant: it is chiral,  meaning that it is covariantly $\bar \theta$-independent, but  it  is required to be also covariantly $ \theta$-independent,  $D^{k}_{ \gamma} \, X=0$, due to torsion in the geometry. It  follows that it cannot  depend on space-time coordinates $x$ due to
 $\{ D_{\a}^i, D_{\dot \b j}\}  X = \delta^i{}_j \partial _{\a \dot \b} X +\dots =0$
 \be
 D_{\dot \a \, i} \, X=0 + {\rm integrability} \qquad  \Rightarrow \quad D^{k}_{ \gamma}\,  X=0 \qquad  \Rightarrow \quad X={\rm const}\ .
 \ee
 
Similarly, if we study the algebra of  nonlinear supersymmetry acting on an $SU(8)$ vector, we find that  two local supersymmetry transformations generate a local $SU(8)$ rotation on an $SU(8)$ vector $X^k$ 
 \be
\{ D_{(\a}^i, D_{\b)}^j \}  X^k= \delta^{(i}_l N_{\alpha\beta} ^{j)k}\, X^l +\dots
\label{al} \ee
 where  the $SU(8)$ curvature $N_{\alpha\beta}^{ij}$,
 \be
 N_{\alpha\beta}^{ij}=- {1\over 72} \epsilon ^{ijklmpqr} \chi_{\alpha klm} \chi_{\beta pqr}\, ,
\label{curv} \ee
  is quadratic in fermions. This term  is absent in the linear supersymmetry algebra. In $\cN=5,6$ analogous expressions for  $U(5)$ and  $U(6)$ $\cH$-symmetry curvatures are obtained by truncation.
  The presence of these and other torsions and  curvatures in the  geometry break at the nonlinear level the constraints which  prove the linear supersymmetry of the linearized CT's.
  
{\it Now back to amplitudes}:   In amplitudes the relevant on shell superfields  (sometimes called super-wave function \cite{Drummond:2008vq}) in $d=4$ depend on $\cN$ Grassmann variables $\eta$'s, see for example \cite{Elvang:2013cua,Freedman:2017zgq}. There are  $2\cN$ supercharges, they  depend on  $\cN$ of $\eta$'s and $\cN$ of $\Big({\partial\over \partial \eta}\Big )$'s for each particle in the process.  In \cite{Elvang:2009wd} the most advanced analysis of $N^{K}MHV$ $n$-point superamplitudes is performed. The manifest linear supersymmetry relates various $n$-point amplitudes with fixed $n$ to each other, the superamplitude comes with the    factor $\delta^{2\cN} (\tilde Q)$. This is an important difference with a nonlinear supersymmetry which relates  amplitudes with different number of points $n$  to each other.

Nonlinear supersymmetry requires the relevant $4\cN$ Grassmann coordinates $\theta_{\a}^i, \bar \theta_{\dot \a\, i}$, $\alpha, \dot \a =1,2, \, i=1,\dots , \cN$  of the superspace \cite{Brink:1979nt, Howe:1981gz}, universal for all particles. The geometric superfields are nonlinear in space-time fields, being related to torsions and curvatures of the superspace. The simplest analogy is in general relativity  where the curvature $R_{\mu\nu\lambda\delta}(h)$ depends on gravitational fields $h_{\mu\nu}$ nonlinearly. For example, the third  component of the  superfield $\chi_{\a ijk} (x, \theta) $ which has a spin 1/2 spinor in the first component, is a Weyl spinor $C_{\a \b \gamma\delta}(x)$. Weyl spinor is related to a nonlinear Riemann-Christoffel tensor and 
\be
D_\a^i D_\b^j  D_\gamma^k \,  \chi_{\delta \, ijk}(x, \theta) |_{\theta=0} = C_{\a \b  \gamma\delta} (x).
\ee
  
  The $\eta$-super-wave function-superfields in amplitudes describe a manifest linear supersymmetry of particle states, it is kind of 1/2 BPS state for MHV amplitudes. 
  In nonlinear superspace geometry the superfields depend on  $4\cN$ $\theta's$ and there is no 1/2 or any other fraction subspace of the whole $4\cN$-dimensional superspace. Therefore predictions of the nonlinear supersymmetry work  ``behind the scenes''   in amplitude computations.

4.  We will show that it is impossible to deform $d=4$ $\cN\geq5$ supergravity action while keeping all the symmetries that the classical  action has, local off shell and global on shell. Namely the deformation of the action leads to inconsistencies with either local Poincar\`e supersymmetry or E7. Moreover, the breaking of E7 before local $\cH$-symmetry  is gauge-fixed leads to breaking of local supersymmetry in the unitary gauge. So, {\it in all cases we find a breaking of local nonlinear supersymmetry which is caused by UV divergence}.

\section{E7, local $\cH$ symmetries,  and unitary gauge in $\cN=5,6,8$}\label{sec:2}

$\cN=5,6,8$  supergravities\footnote{The  review of $\cN\geq 5$ supergravities with the proof of absence of $U(1)$ anomalous amplitudes can be found in \cite{Freedman:2017zgq} .} have global duality symmetries, in addition to local symmetries, which complicates the analysis of UV divergences. These are symmetries defined by the groups 
\be
{\cal G} : S U(1,5) , \, \, S O^*( 1 2 ) , \, \, E_{ 7 ( 7 )}
\ee 
 in $\cN=5,6,8$, respectively. These are called groups of type E7, see for example \cite{Ferrara:2011dz} and references therein.  The local symmetries, in addition to local supersymmetry, include local $\cH$-symmetries   
\be
\cH  :  U(5), \, \, U(6),  \, \, SU(8) 
\ee
 in $\cN=5,6,8$, respectively. The scalars in these theories before local $\cH$-symmetries are gauge-fixed  are in the fundamental representation of $\cG$. When local $\cH$-symmetries are gauge-fixed only physical scalars remain. These physical scalars represent the coordinates of the coset space ${\cG\over \cH}$. For example, in $\cN=8$ there are 133 scalars before local $SU(8)$ is gauge-fixed, and only 70 physical scalars in the unitary gauge, the 63 local parameters of $SU(8)$ being used to remove the unphysical scalars.
 
The vector fields transform as  doublets under E7, however, only half of them  are  physical vectors. The relevant constraint on graviphotons takes care of the unitarity of the theory, making  half of the doublets to be physical and independent, the second half dependent on physical vectors.  For example, in $\cN=8$ there are 56 vectors in the doublet but only 28 of them are physical. Therefore E7 duality and the related self-duality constraint are required for the unitarity of the theory.

A constraint which makes  half of supergravity vectors  physical was introduced in~\cite{Cremmer:1979up,deWit:1982bul}. It was given a name {\it twisted nonlinear self-duality constraint}  and studied   in \cite{Cremmer:1998px,Gaillard:1981rj,Andrianopoli:1996ve,Hillmann:2009zf,Bossard:2010dq, Kallosh:2011dp,Kallosh:2011qt,Bossard:2011ij,Carrasco:2011jv,Pasti:2012wv,Kallosh:2018mlw}. 
Thus when looking at the deformation of the action caused by potential UV divergences we need to preserve local supersymmetry, local $\cH$-symmetry and E7 symmetry. All these symmetries may require  a deformation consistent with a deformed action. 

The \E\  duality in $\cN=8$ was discovered and studied in \cite{Cremmer:1979up,deWit:1982bul}  where 133 scalars are present before the gauge-fixing of a local $SU(8)$ symmetry. The gauge-fixing  of local $SU(8)$ was also performed in  \cite{Cremmer:1979up,deWit:1982bul,Kallosh:2008ic} and the unitary gauge with 70 physical scalars was described.

A general case of dualities in $d=4$ supergravities was introduced  by Gaillard and Zumino (GZ) in \cite{Gaillard:1981rj}.  Standard global symmetries require a Noether current conservation, but in case of GZ duality  \cite{Gaillard:1981rj} the usual Noether procedure  is not applicable since duality acts on field strength and its dual rather than on vector fields.  Therefore this duality symmetry is associated with the Noether-Gaillard-Zumino (NGZ) current conservation. In $\cN=8$ case this NGZ conserved current  was presented in \cite{Kallosh:2008ic}.

Studies of duality symmetries  were also performed  in a unitary gauge,  where local $\cH$-symmetries of supergravities were gauge-fixed \cite{Cremmer:1979up,deWit:1982bul,Kallosh:2008ic},  or using a symplectic formalism~\cite{Andrianopoli:1996ve} developed in the bosonic theory without fermions. In both cases  only physical scalars are present in the theory \cite{Gaillard:1981rj,Andrianopoli:1996ve,
Hillmann:2009zf,Bossard:2010dq,Broedel:2009nsh,Elvang:2010jv,Beisert:2010jx,Freedman:2011uc,Kallosh:2011dp,Kallosh:2011qt,Bossard:2011ij,Carrasco:2011jv,Pasti:2012wv,Kallosh:2018mlw}, they form coordinates of the coset space ${\cG\over \cH}$.

The approach to deformation of  $\N\geq 5$ supergravity developed in \cite{Bossard:2011ij,Kallosh:2018mlw} was 
revisited in \cite{Kallosh:2012yy,Gunaydin:2013pma, Kallosh:2018wzz,Gunaydin:2018kdz} from the point of view of special properties of E7-type groups and  supersymmetry.
  It was shown there that in absence of supersymmetry anomalies, duality symmetry protects $\cN\geq 5$ supergravities from UV divergences. In \cite{Kallosh:2018wzz} the analysis was based on manifest E7 symmetry whereas  that
 in \cite{Gunaydin:2018kdz} was based on the properties of the unitary conformal supermultiplets of $SU(2,2|\cN+n)$. 
 
 Here we will first recall that at the loop order $L<L_{cr} = \cN$ for $\cN=5,6,8$ there are no geometric superinvariants \footnote{ In $\cN=4$ the situation is different since the version of the theory with unbroken local $U(4)$ symmetry has also a local superconformal symmetry, see \cite{Ferrara:2012ui} and references therein. Local superconformal symmetry in $\cN=5,6,8$ supergravities is broken 
  at the nonlinear level.}
 in whole $(x, \theta)$ superspace  in $d=4$.  This means that
  a UV divergence at $L<L_{\rm cr} = \cN$  breaks nonlinear local supersymmetry and local $\cH$ symmetry. If such terms are added to the classical action, they will break nonlinear local supersymmetry and local $\cH$ symmetry of the classical theory. It means that the classical  action deformed by $L<L_{\rm cr} = \cN$ CT's is  BRST inconsistent since deformed action is not invariant under local symmetries of the classical action.
  
We will also study the deformation of the local supersymmetry transformation caused by potential UV divergences supported by geometric non-linearly supersymmetric and $\cH$ locally invariant candidate CTs at $L\geq \cN$. We  will look at  {\it supersymmetry transformation of fermions before gauge-fixing where fermions  transform under
local  $\cH$-symmetry, they are neutral under E7, and these two symmetries are independent and linearly realized}.

 Fermions  before gauge-fixing local $\cH$-symmetry transform under supersymmetry into an E7 invariant and $\cH$-covariant graviphoton $\cF$, for example
\be
\delta_S \chi_{ijk} = \cF_{ij} \epsilon _k +\dots\ .
\ee
In presence of the CT deforming the action  the graviphoton is deformed into $\cF_{ij}^{\rm def}$
\be
\cF_{ij}^{\rm def} = \cF_{ij}^{\rm (cl)} + \l \hat \cF_{ij}.
\ee
This requires a deformation of supersymmetry transformation on fermions 
\be
\delta^{\rm def}_S \chi_{ijk} = \cF_{ij}^{\rm def} \epsilon _k +\dots=\cF_{ij}^{\rm (cl)}\epsilon_k+ \l \hat \cF_{ij}\epsilon_k+\cdots.
\label{defFer}\ee 
to preserve the invariance  of the fermions on E7  symmetry and covariance on $\cH$-symmetry before gauge-fixing. We will find out that the deformed action \rf{Def} is not invariant under deformed supersymmetry~\rf{defFer}.

If, instead, we sacrifice E7 and do not deform supersymmetry transformation on fermions, we will find that the breaking of E7 before gauge-fixing feedbacks into  breaking local nonlinear supersymmetry in the unitary gauge.

\section{Symmetries of  de Wit-Nicolai (dWN) $\cN=8$ supergravity}\label{sec:3}
 \subsection{Local $SU(8)$ and on shell \E\, symmetry }
 
 The classical action \cite{deWit:1982bul} with a local $SU(8)$ symmetry before this symmetry is gauge fixed   depends on 133 scalars represented by a  56-bein
\begin{eqnarray}\label{gauge}
{\cal V}=\left(
                                        \begin{array}{cc}
                                          u_{ij} {} ^{IJ}& v_{ijKL} \\
                                           v^{klIJ} &  u^{kl}{}_{KL} \\
                                        \end{array}
                                      \right)\                                      \end{eqnarray}
                                      and its inverse
\begin{eqnarray}\label{in}
 {\cal V}^{-1}=\left(
                                        \begin{array}{cc}
                                          u^{ij} {} _{IJ}& -v^{ijKL} \\
                                          - v_{klIJ} &  u_{kl}{}^{KL} \\
                                        \end{array}
                                      \right)\ .
                                      \end{eqnarray}                                      
 We summarize some identities for these matrices in appendix~\ref{appB}.                                    
The capital indices $I,J$ refer to \E\,  and small  ones $ij$ refer to $SU(8)$. The 56-bein transforms under a local $SU(8)$ symmetry $U(x)$ and a global \E\,  symmetry    $E $   as follows
\be
{\cal V}(x) \rightarrow U(x) {\cal V}(x) E^{-1}.
\label{Vtransf}\ee    
These two symmetries are linearly realized and independent. Here $E \in E_{7(7)}$    is in the fundamental 56-dimensional representation where
\begin{eqnarray}\label{E7}
E=\exp \left(
                                        \begin{array}{cc}
                                          \Lambda_{IJ} {} ^{KL}& \Sigma _{IJPQ} \\
                                          \Sigma ^{MNKL} &  \Lambda ^{MN}{}_{PQ} \\
                                        \end{array}
                                      \right)\ .
                                      \end{eqnarray}
Duality symmetry in \rf{E7} consists of a diagonal transformation $\Lambda _{IJ}{}^{KL}= \delta _{[I}{}^{K} \Lambda _{J]}{}^{L} $ where $\Lambda _{J}{}^{L}$ are the generators of the $SU(8)$ maximal subgroup of \E\ with 63 parameters. The off-diagonal part is  self-dual $\Sigma _{IJPQ}= \pm {1\over 4!} \epsilon_{IJPQ MNKL}   \Sigma ^{MNKL}$ and has 70 real parameters.

 The total Lagrangian in eq. (3.18) of \cite{deWit:1982bul} consists of two parts. One is manifestly \E$\times SU(8)$ invariant, where \E\ is a global symmetry and  $SU(8)$ is a local symmetry. The other part of the action $\cL' + \cL''$ is  not manifestly  \E$\times SU(8)$ invariant. It depends on vectors, scalars and fermions  and  takes 3 lines in eq.  (3.18) in  \cite{deWit:1982bul} (lines 3, 4, 5).
 
 The 28 abelian vector field strength are defined   as 
 \be
 F^{IJ} _{\mu\nu } =\partial_\mu  A_\nu ^{IJ}- \partial_\nu  A_\mu ^{IJ}= F^+_{\mu\nu IJ} + F^{-IJ} _{\mu\nu }\, .
 \ee
The dual vector field strength is
\be
G^{+\mu\nu}_{IJ} \equiv -{4\over e} {\partial {\cal L}\over \delta F^+_{\mu\nu IJ} }.
\label{G}\ee
The same Lagrangian $\cL' + \cL''$ takes the following  form in these notations  
 \bea
\cL' + \cL'' =  -{1\over 8} e F^+_{\mu\nu IJ} \,  G^{+\mu\nu}_{ IJ}  
- {1\over 4} e \cF^+_{\mu\nu ij} \,  \cO^{+\mu\nu ij} +{\rm h.c.}\ .
\label{sym} \eea
 Here 
 the fermion bilinear term is
 \bea
{\cal O} ^{+ ij}_{ \mu\nu} = \pm \Big [ {1\over 144} \sqrt 2  \epsilon ^{ijklmnpq} \bar \chi_{klm} \sigma_{\mu\nu} \chi_{npq} -{1\over 2} (\bar \psi_{\lambda k} \sigma_{\mu\nu} \gamma^\lambda \chi^{ijk} - \sqrt 2 \bar \psi^i _\rho \gamma^{[\rho} \sigma_{\mu\nu} \gamma^{\sigma]} \psi^j_\sigma) \Big ].
\label{Oij} \eea  
The first term is bilinear in gaugino, the second one has  gaugino and a gravitino, the third one is bilinear in gravitino. 
   
  The graviphoton here is related to $F^+_{\mu\nu KL}$ as~\cite{deWit:1982bul}
\be
u^{ij}{}_{IJ} {\cal F}^{+}_{\mu\nu ij}  = S^{IJ,KL} F^{+}_{ \mu\nu KL} + (S^{IJ,KL}+u^{ij}{}_{IJ}v_{ijKL}) {\cal O}_{\mu\nu} ^{+ KL},
\label{graviphotonCL}\ee
where $S^{IJ,KL}$ is defined in eq.~\eqref{Sdefinition}. We will review the derivation of this relation below.

Now the first term in the action in \rf{sym} together with its h.c. vanishes on shell since upon partial integration it is of the form
\be
F\tilde G \rightarrow A_\nu \partial _\mu \tilde G^{\mu\nu}|_{\rm on \ shell} =0
\ee
and the  vector equation of motion with account of \rf{G} is
\be
\partial _\mu \tilde G^{\mu\nu}=0 \, .
\label{Geq}\ee
This equation of motion   pairs with the Bianchi Identity for the abelian vector field strength
\be
\partial _\mu \tilde F^{\mu\nu}=0\, .
\ee
E7-type symmetry flips one into another. The second term in  \rf{sym}  depends on the graviphoton $\cF^+_{\mu\nu ij} $,    and on spinor bilinear  $ \cO^{+\mu\nu ij}$ which are both  $SU(8)$ tensors and E7 invariants. 

To conclude, the 3-line part of the action in  \cite{deWit:1982bul} can be brought to a form \rf{sym} which, on shell with account of eq. \rf{Geq},  has manifest local $SU(8)$ symmetry and global \E\ .

\subsection{Manifest \E\, and twisted nonlinear self-duality constraint}

The  
  \E\, doublet $\left( \begin{array}{c}
                                          F^{+}_{1 \mu\nu IJ}  \\
                                          F^{+ IJ}_{2 \mu\nu}  \\
                                        \end{array}
                                      \right)$ depends on a double set of vectors, 56 in this case,  which is a minimal representation for the symplectic representation in \E\,. Note however that there are only 28 physical vectors in $\cN=8$ supergravity. The doublet   is constructed in \cite{deWit:1982bul} from a combination of the field strength $F$ and its dual $G$, where $G$ defined in eq. \rf{G}
                                      \be
                                       F^{+}_{1 \mu\nu IJ} \equiv  {1\over 2} (G^{+\mu\nu}_{IJ} + F^{+\mu\nu}_{IJ}),
                                      \label{F1}\ee
                                      \be
                                       F^{+ IJ}_{2 \mu\nu} \equiv  {1\over 2} (G^{+\mu\nu}_{IJ} - F^{+\mu\nu}_{IJ}).
                                       \label{F2}\ee
The  \E\, doublet $\left( \begin{array}{c}
                                          F^{+}_{1 \mu\nu IJ}  \\
                                          F^{+ IJ}_{2 \mu\nu}  \\
                                        \end{array}
                                      \right)$ depends on 56 independent vectors which is necessary to have manifest \E\,.   Under \E\, the doublet transforms as 
   \be\left( \begin{array}{c}
                                          F^{+}_{1 \mu\nu IJ}  \\
                                          F^{+ IJ}_{2 \mu\nu}  \\
                                        \end{array}
                                      \right)\rightarrow E \left( \begin{array}{c}
                                          F^{+}_{1 \mu\nu IJ}  \\
                                          F^{+ IJ}_{2 \mu\nu}  \\
                                        \end{array}
                                      \right).\label{E7doublet}
                                      \ee    
                                      We define the $SU(8)$ covariant graviphoton and ${\cal F}^{+}_{\mu\nu ij}$ and  the $SU(8)$ covariant tensor ${\cal T}_{\mu\nu}^{+ij}$
\begin{equation}
\left(\begin{array}{c}{\cal F}^{+}_{\mu\nu ij}\\ {\cal T}_{\mu\nu}^{+ij}\end{array}\right)\equiv {\cal V} \left(\begin{array}{c}F^+_1\\ F^+_2\end{array}\right)=\left(\begin{array}{cc}u_{ij}{}^{IJ}&v_{ijKL}\\ v^{ijIJ} & u^{ij}{}_{KL}\end{array}\right)\left(\begin{array}{c}F^+_{1\mu\nu IJ}\\ F^{+KL}_{2\mu\nu}\end{array}\right) .\label{E7inv}          
\end{equation}                
  All capital indices in the r.h.s. of eq.   \rf{E7inv}   are contracted between the  56-bein  ${\cal V}$ in eq. \rf{gauge} and E7 doublet
From their definition and \E\, properties of $\cV$ and the doublet  $\left( \begin{array}{c}
                                          F^{+}_{1 \mu\nu IJ}  \\
                                          F^{+ IJ}_{2 \mu\nu}  \\
                                        \end{array}
                                      \right)$ it is clear that the l.h.s. of eq. \rf{E7inv}  are \E\, invariant. Thus, both the graviphoton and the tensor ${\cal T}_{\mu\nu}^{+ij}$ in eq. \rf{E7inv}     are manifestly invariant under \E\, symmetry according to \rf{Vtransf} and  \rf{E7doublet}.  
                                      
In this manifestly E7 invariant form the constraint on vectors which makes only half of them physical and the other half dependent on physical vectors and scalars takes the form  
\begin{equation}
{\cal T}_{\mu\nu}^{+ij}={\mathcal O}^{ij}_{\mu\nu},\label{classicalTDC}
\end{equation}
or equivalently
\begin{eqnarray} 
{1\over 2} (I-\Omega) {\cal V} \left( \begin{array}{c}
                                          F^{+}_{1 \mu\nu }  \\
                                          F^{+ }_{2 \mu\nu}  \\
                                        \end{array}
                                      \right)=
\left( \begin{array}{c}
                                          0  \\
                                          {\cal O} ^{+ ij}_{ \mu\nu}  \\
                                        \end{array}
                                      \right).
\label{con} \end{eqnarray}    
Here the 56-dimensional $\Omega$ is
\be
\Omega = \left(
                                        \begin{array}{cc}
                                          I& 0 \\
                                         0 &  -I \\
                                        \end{array}
                                      \right)\   .
\ee   

We call eq.~\eqref{classicalTDC} {\it the twisted self-duality constraint}. Let us see how eq.~\eqref{classicalTDC} reduces the number of degrees of freedom: Equation.~\eqref{classicalTDC} is explicitly given by
\begin{equation}
2{\cal O}^{+ij}=(v^{ijIJ}+u^{ij}{}_{IJ})G^{+}_{IJ}+(v^{ijIJ}-u^{ij}{}_{IJ})F^{+IJ}.
\end{equation}
Note that here and hereafter we omit the Lorentz indices. We need to solve it with respect to the dual field strength $G_{IJ}^+$, which yields
\begin{align}
G^+_{IJ}=&(v^{ijIJ}+u^{ij}{}_{IJ})^{-1}\left[(u^{ij}{}_{KL}-v^{ijKL})F^{+KL}+2{\cal O}^{+ij}\right]\nonumber\\
=&\left[(2S-\mathbf{1})^{IJ,KL}F^{+KL}+2S^{IJ,KL}{\cal O}^{+KL}\right],
\end{align}
where ${\cal O}_{\mu\nu}^{+ij}\equiv u^{ij}{}_{IJ}{\cal O}^{+IJ}_{\mu \nu}$ and we have used some identities in appendix~\ref{appA}. This result corresponds to (2.4) of \cite{deWit:1982bul}. One can rewrite graviphoton ${\cal F}_{ij}^+$ as
\begin{align}
{\cal F}^{+\rm (cl)}_{ij}=&\frac12 (u_{ij}{}^{IJ}+v_{ijIJ})G_{IJ}^++\frac12 (u_{ij}{}^{IJ}-v_{ijIJ})F^{+}_{IJ}\nonumber\\
=&\frac12(u_{ij}{}^{IJ}+v_{ijIJ}) \left[2S^{IJ,KL}F^+_{KL}-F_{IJ}^++2S^{IJ,KL}{\cal O}_{KL}^+\right]+\frac12 (u_{ij}{}^{IJ}-v_{ijIJ})F_{IJ}^+\nonumber\\
=&(u_{ij}{}^{IJ}+v_{ijIJ})S^{IJ,KL}F_{KL}^+-v_{ijIJ}F_{IJ}^++(u_{ij}{}^{IJ}+v_{ijIJ})S^{IJ,KL}{\cal O}_{KL}^+\nonumber\\
=&({\cal M}_{ij,kl}u^{kl}{}_{KL}-v_{ijKL})F_{KL}^++{\cal M}_{ij,kl}{\cal O}^{+kl},\label{CLgraviphoton}
\end{align}
where 
\be
{\cal M}_{ij,kl}\equiv (u_{ij}{}^{IJ}+v_{ijIJ})(u^{kl}{}_{IJ}+v^{klIJ})^{-1}.\label{calM}\ee 
The twisted self-duality constraint \eqref{classicalTDC} reduces the number of degrees of freedom in a manifestly $E_{7(7)}$ invariant way.

In conclusion, the classical action of $\cN=8$ supergravity in the form \rf{sym} has the following properties. The first term is
\be
F \tilde G 
\ee
and it vanishes on shell when classical field equations are satisfied. The second term involves a graviphoton coupled to fermion bilinears
 \be
\cL_{\rm cl} =  
- {1\over 4} e \cF^+_{\mu\nu ij} \,  \cO^{+\mu\nu ij} +{\rm h.c.} + \dots \ .
\label{sym1} \ee
Here $\dots$ include vector independent terms, which are manifestly \E\, and local $SU(8)$ invariant.

Once the CT  is added to the classical action, equations of motion are deformed. We will study this below for $\cN\geq 5$ supergravities in general, and provide details in $\cN=8$ case.

\section{ dWN supergravity deformation}\label{sec:5}
The deformation of Lagrangian due to the presence of a new local CT must be consistent with the duality, and therefore the dual field strength is affected by the presence of the new term in the action so that $G\rightarrow G^{\rm def} $
\be
G^{+\, \rm def }_{KL} = -4   {\delta (\cL^{\rm cl}   +\l {\cal L}^{\rm CT})\over \delta F^+_{\mu\nu KL} }.\label{Gdefdef}
\ee
In order to keep manifest \E\, and  $SU(8)$ invariance, instead of starting from the action, we deform the twisted duality condition~\eqref{classicalTDC} and find a consistent dual field strength and a corresponding action, which is proposed in \cite{Bossard:2011ij}. Here we generalize the previous results~\cite{Bossard:2011ij, Kallosh:2018mlw} to include fermionic bilinear terms: We consider the deformed twisted self-duality constraint
\begin{equation}
{\cal T}_{\mu\nu}^{+ij}+\lambda X^{ij}{}_{kl}\bar{\cal F}^{-kl}_{\mu\nu}={\mathcal O}^{+ij}_{\mu\nu}\label{defTDC}
\end{equation}
where $X^{ij}{}_{kl}$ is an ${\cal H}$-covariant differential operator depending on other fields such as scalars and gravitons. 
Since we may interpret this condition as a shift of ${\cal O}^{+ij}_{\rm def}={\cal O}^{+ij}-\lambda X^{ij}{}_{kl}\bar{\cal F}^{-kl}$, one can formally rewrite \eqref{defTDC} as
\begin{align}
{\cal F}^+_{ij}={\cal F}^{+\rm(cl)}_{ij}-\lambda {\cal M}_{ij,kl}X^{kl}{}_{mn}\bar{\cal F}^{-mn}
\label{dg}\end{align}
or equivalently
\begin{equation}
\bar{\cal F}^{-ij}=\bar{\cal F}^{-ij\rm (cl)}-\lambda\bar{\cal M}^{ij,kl}\bar{X}_{kl}{}^{mn}{\cal F}^{+}_{mn},
\end{equation}
where the classical graviphoton is defined in \eqref{CLgraviphoton} and we recall that ${\cal M}_{ij,kl}\equiv (u_{ij}{}^{IJ}+v_{ijIJ})(u^{kl}{}_{IJ}+v^{klIJ})^{-1}$.
One can substitute the second equation to the first, which yields
\begin{align}
&{\cal F}^+_{ij}={\cal F}^{+\rm(cl)}_{ij}-\lambda {\cal M}_{ij,kl}X^{kl}{}_{mn}(\bar{\cal F}^{-mn\rm (cl)}-\lambda\bar{\cal M}^{mn,pq}\bar{X}_{pq}{}^{rs}{\cal F}^{+}_{rs})\nonumber\\
\Leftrightarrow & ({\bf 1}-\lambda^2{\cal M}X\bar{\cal M}\bar{X})_{ij}{}^{kl}{\cal F}_{kl}^+={\cal F}^{+\rm(cl)}_{ij}+\lambda{\cal M}_{ij,kl}X^{kl}{}_{mn}\bar{\cal F}^{-mn\rm (cl)}\nonumber\\
\Leftrightarrow& {\cal F}_{kl}^+=\left(({\bf 1}-\lambda^2{\cal M}X\bar{\cal M}\bar{X})^{-1}\right){}^{kl}{}_{ij}({\cal F}^{+\rm(cl)}_{kl}+\lambda{\cal M}_{kl,mn}X^{mn}{}_{pq}\bar{\cal F}^{-pq\rm (cl)}).
\label{allorder}\end{align}
This is a formal all-order solution to the deformed twisted self-duality condition. We emphasize that in the derivation of the deformed graviphoton, we have not imposed gauge fixing conditions on local $SU({\cal N})$ and therefore the result is fully consistent with both ${\cal G}$ and ${\cal H}$.  One can further solve this relation in $G_{IJ}^+$, which fixes the dual field strength as the classical supergravity case.

Formal expansion in $\lambda$ yields the graviphoton, and up to ${\cal O}(\lambda)$ we find
\begin{equation}
{\cal F}_{ij}^+={\cal F}^{+\rm(cl)}_{ij}+\lambda{\cal M}_{ij,kl}X^{kl}{}_{mn}\bar{\cal F}^{-mn\rm (cl)}+{\cal O}(\lambda^2).\label{graviphotondef}
\end{equation}
Solving this equation with respect to $G_{KL}^{+\rm def}$ and integrating the both sides of \eqref{Gdefdef} with respect to $F^+_{KL}$  perturbatively yields the corresponding duality invariant action, and one can check that the lowest order correction is given by
\begin{equation}
\lambda \, {\cal L}^{CT}=- {1\over 2}\lambda \,  {\cal F}^{ +}_{  ij }  X^{ij}{}_{kl}  \bar  {\cal F}^{ - kl } +{\cal O}(\lambda^2)\label{pertCT}
\end{equation}
as we expected. However, the constructed action in general has infinite numbers of higher order terms, which is necessary to keep $E_{7(7)}$ to all orders. This is a generalization of the purely bosonic deformation in \cite{Kallosh:2011qt,Bossard:2011ij,Carrasco:2011jv,Kallosh:2018mlw}, and we also emphasize that we have not gauge fixed local $\cal H$-symmetry unlike our previous construction using a symplectic formalism in  \cite{Andrianopoli:1996ve}.

We would like to emphasize the most crucial point of our result~\eqref{graviphotondef}: The on-shell deformed graviphoton has a term including $M_{ij,kl}$ which looks $E_{7(7)}$ invariant but actually is not. Thus, the deformation of the graviphoton makes the $E_{7(7)}$ invariance not manifest. Nevertheless, it is still consistent with $E_{7(7)}$ if we include all order corrections by construction. However, it is not clear if such deformation is consistent with supersymmetry, and we will show that it is unlikely, from which we will conclude that despite duality invariant construction of the deformation, the resultant action leads to a problem with supersymmtry.

Throughout this paper, we focus on the deformed twisted self-duality constraint~\eqref{defTDC}. One may wonder whether a more general constraint is available. We expect that it is possible  but the generalization would not change our conclusion: We note that $\lambda$ is a coupling constant which is some powers of $\kappa$, and we could add more terms to the constraint such as $\lambda({\cal F}^{+\rho \eta }_{kl}{\cal T}_{\rho\eta}^{+kl})^n{\cal T}^{+ij}_{\mu\nu}$ if $n$ is appropriately chosen, namely, if the mass dimension of $\lambda({\cal F}^{+\rho \eta }_{kl}{\cal T}_{\rho\eta}^{+kl})^n{\cal T}^{+ij}_{\mu\nu}$ matches that of $\lambda X^{ij}{}_{kl}\bar{\cal F}^{-kl}_{\mu\nu}$. Such terms would contribute to higher point interactions and we will not discuss it here as we are interested in a minimal deformation.\footnote{We expect that such deformation can also be solved at least perturbatively in $\lambda$.} As far as we have considered, we have not found any term that (1) may change our discussion below, (2) has appropriate $SU(8)$ indices and (3) is manifestly $E_{7(7)}$ invariant. Therefore, we believe that the following discussion would not be changed by adding more terms to the twisted self-duality constraint.

\section{Deformation of $\cN\geq 5$ }\label{sec:4}

\subsection{Candidate CT's}\label{sec:40}
There are three  approaches to candidate CT's we would like to  describe here shortly.

1. The candidate CT's for the possible UV divergences
in extended supergravities 
 were predicted in the past \cite{Kallosh:1980fi,Howe:1980th} on the basis of a Lorentz-covariant on shell superspace geometry \cite{Brink:1979nt, Howe:1981gz} with 4 space-time coordinates $x$ and $4\cN$ Grassmann coordinates $\theta$. These were either linearized CT's or full nonlinear CT's, the examples will be presented below. The nonlinear CT's are known to have manifest local nonlinear supersymmetry and duality symmetry under condition that classical equations of motion are satisfied. 
 
 Linearized CT's break nonlinear supersymmetry and duality, some of them can be promoted to full nonlinear status, some cannot. The difference is defined by dimension: In $d=4$ the ones for loop order $L\leq \cN -1 $ cannot be promoted to the supersymmetric terms at the nonlinear level, whereas the ones for $L\geq  \cN  $ can be.
 
2. The candidate CT's in Lorentz-covariant on shell harmonic superspace geometry, see for example  \cite{Bossard:2011tq}, are linearized CT's depending on additional harmonic coordinates. At the linearized level they can be written also without harmonic coordinates as integrals over subspace of the superspace $\int d^{4\cN(1-{1\over k})}\theta$, these are called ${1\over k}$ BPS invariants. We have argued in Sec.~\ref{sec:12} that nonlinear harmonic superspace describing  nonlinear super-Poincar\'e supergravities is inconsistent since the relevant constraints promised in \cite{Hartwell:1994rp} is still missing.

3. Finally  candidate CT's were studied in the amplitude's framework in \cite{Elvang:2010jv,Beisert:2010jx,Freedman:2011uc,Freedman:2018mrv}. In all cases in these papers only linearized supersymmetry was used in combination with studies of a single soft scalar limit. In all cases in \cite{Beisert:2010jx,Freedman:2018mrv}  only part of nonlinear supersymmetry and duality was used, namely linearized supersymmetry and soft scalar limits. This was the reason why  in \cite{Beisert:2010jx} the case of $\cN=8, L=7$ was left as inconclusive and the same in \cite{Freedman:2018mrv}  where the case of $\cN=5, L=4$ was left as inconclusive.

By comparing these tree  approaches to candidate CT's we conclude that only in case 1. we have a clear explanation of enhanced cancellation of $\cN=5, L=4$ UV divergence \cite{Bern:2014sna} since the relevant candidate CT breaks nonlinear local supersymmetry.

We assume that UV divergences require the deformation of the action so that we add the CT with the parameter $\l$ 
\be
S^{\rm def} = S_{\rm cl} +\l S^{\rm CT}_{L\leq \cN-1}.
\ee
In what follows we will use the short form of the superinvariants as integrals over a superspace or its subspaces. Note that in our eqs. \rf{CT8}, \rf{L3}, \rf{CT8L} below we show that the result of $\theta$-integration of these superinvariants can be computed and gives a space-time integral with some dependence on space time curvature as well as other terms shown by a set of dots.

The reason to use a short form of the supersymmetric invariants becomes clear if one looks at the 3-loop CT in $\cN=8$ in eq. (6.8) in \cite{Freedman:2011uc} written in components in linearized approximation: it has 51 terms. This expression was obtained  using  amplitude methods. But all these 51 terms  are  packaged in the linearized superinvariant in eq. \rf{L3} below, it was first  presented in \cite{Kallosh:1980fi}.

\subsection{Loop order $L\leq \cN -1 $}\label{sec:41}
It is known 
from \cite{Kallosh:1980fi,Howe:1980th}  that in $d=4$ the whole superspace CT's are available only starting from $L_{\rm cr} =\cN$\be
CT^{L=\cN}=\kappa^{2(\cN-1)} \int d^4 x \, d^{4\cN} \theta  \det E \,  \cL (x, \theta)  = \kappa^{2(\cN-1)} \int d^4 x \, D^{2(\cN-3)} \, R^4 +\dots
\label{CT8}\ee
where the superspace Lagrangian $\cL (x, \theta)$ has dimension 2,  the smallest possible dimension for a geometric Lagrangian. For example at $\cN=8$ it is a  quartic product  of 4  geometric superfields defining minimal dimension  torsion shown in  eq. \rf{torsion}. 
\be
\cL (x, \theta)^{L=8}=   \chi_{\alpha \, ijk} (x, \theta)\,  \chi_{mnl}^\alpha (x, \theta) \,  \bar \chi_{\dot \alpha} ^{ ijk} (x, \theta)\, \bar \chi^{ \dot \alpha \, mnl}  (x, \theta) \label{L8}\ee 
The first components of these superfields are spin 1/2 fermions.
 At smaller $\cN$ a consistent supersymmetric truncation of the maximal superspace \cite{Brink:1979nt} was performed in  \cite{Howe:1981gz}.  A consistent truncation of the expression in \rf{L8} will provide the superfield Lagrangian for smaller $\cN$. 
Here we integrate over the total superspace and the spinorial superfield $\chi_{\alpha \, ijk}$  associated with the superspace torsion is covariant under the local $SU(8)$. The superspace Lagrangian \rf{L8} quartic in spinorial superfields is invariant under local $\cH$-symmetry.

All candidate CT's at $L<L_{\rm cr}$ are available only as integrals over a subspace of the superspace: this is one way of reducing the dimension of  $d^{4\cN} \theta$. Also the superfield Lagrangian in linearized supersymmetry is not geometric anymore, it typically depends on superfields starting with scalar fields and has dimension 0, instead of dimension 2 in \rf{L8}.

For example, the 3-loop $\cN=8$ candidate CT is \cite{Kallosh:1980fi,Bossard:2009sy}
\be
CT^{L=3}=\kappa^{4} \int d^4 x \, (d^{16} \theta)_{1234}  W^4_{1234} = \kappa^{4} \int d^4 x \,  R^4 +\dots \ .
\label{L3}\ee
It is an integral over the half of the superspace and it depends on physical scalars only. This linearly supersymmetric expression exist only in the unitary gauge where the local $SU(8)$ symmetry is gauge-fixed. In loops $L=4,5,6,7$ the linearized candidate CT's are also available, and they also require a unitary gauge and a subspace of the full superspace. Therefore all local symmetries, in particular a nonlinear local supersymmetry is broken, despite the CT's like \rf{L3} have unbroken linear supersymmetry. It means, for example, that the simplest explanation of the 3-loop UV finiteness of $\cN=8$  supergravity \cite{Bern:1998ug} is the fact that the CT \rf{L3} breaks nonlinear local supersymmetry.

 If a UV divergence will show up at $L<L_{\rm cr}= \cN$ in in $\cN=5,6,8$ and  the corresponding CT will be added to the action to absorb the UV divergence,  the relevant deformed theory will be BRST inconsistent since a local nonlinear supersymmetry of the deformed action will be broken. There is even no need to study the situation with duality in these cases, breaking of local symmetries  makes the deformed action BRST inconsistent. 

So far in the loop computations in $d=4$ we have not seen UV divergences  at $L<L_{\rm cr}= \cN$. The loop computation in $\cN=5, L=4$ \cite{Bern:2014sna} suggest that so far there is no need to deform $\cN=5$ supergravity. But the loop computations of  UV divergences at $\cN=6, L=5$ and $\cN=8, L=7$ are not available. 
  
  Thus we have to wait to see if $d=4$ is special in this respect. Assuming that as in $\cN=5$ case the cases of $\cN=6, L=5$ and $\cN=8, L=7$ are also UV finite, we proceed to the case $L\geq \cN  $ for all of them.

\subsection{Loop order $L\geq \cN  $}\label{sec:42}
Starting from loop order $L=\cN$  the geometric on shell CT's are available \cite{Kallosh:1980fi,Howe:1980th}. With a symbolic insertion of space-time $\cH$ covariant  derivatives to increase the dimension we can present them as follows
\be
CT^{L\geq \cN}=\kappa^{2(L-1)} \int d^4 x \, d^{4\cN} \theta  \det E \,  \cL (x, \theta)  = \kappa^{2(L-1)} \int d^4 x \, D^{2(L-3)} \, R^4 +\dots,
\label{CT8L}\ee
\be
\cL (x, \theta)=   \, \chi_{\alpha \, ijk} (x, \theta)\,  \chi_{mnl}^\alpha (x, \theta) \, D^{2(L-\cN)}  \bar \chi_{\dot \alpha} ^{ ijk} (x, \theta)\, \bar \chi^{ \dot \alpha \, mnl}  (x, \theta), \label{L}\ee 
where $D$ in \eqref{CT8} denotes spacetime  covariant derivative whereas that in \eqref{L} symbolically denotes multiples of either spinor or spacetime covariant and $\cH$-covariant derivatives with total dimension $2(L-\cN)$.   
These expressions require that the classical equations of motion are valid since the superspace in $\cN\geq 5$ is available only on shell  \cite{Brink:1979nt, Howe:1981gz}.

At this point if the UV divergence takes place at any of $L\geq \cN  $,  we deform the action to absorb UV divergence, as 
\be
S^{\rm def} = S_{\rm cl} +\l S^{\rm CT}_{L\geq \cN}.
\ee
We cannot easily dismiss these terms  as in cases $L\leq \cN -1$ where the CT's like the ones in \rf{N5} and in \rf{L3} manifestly break non-linear supersymmetry and local $\cH$ symmetry and therefore do not present a  consistent deformation. As long as classical equations of motion are satisfied the CT's in eqs. \rf{CT8}, \rf{L} appear to be legitimate candidates for the deformation.

However, once we deform the classical action due to UV divergences, classical equations of motion are not valid anymore, they acquire $\l$-corrections. Furthermore, it was already realized that due to these corrections, E7 symmetry of the deformed action is broken and higher order in $\l$ deformations are required to restore E7 symmetry \cite{Kallosh:2011qt,Bossard:2011ij,Carrasco:2011jv,Kallosh:2018mlw}. The study in \cite{Kallosh:2018mlw} was performed  for the bosonic action with  local $\cH$-symmetry gauge-fixed.

Here we have generalized the results in \cite{Kallosh:2018mlw} to the supergravity with local $\cH$-symmetry and including fermions in Sec. \ref{sec:5}. We will study below the local supersymmetry of the deformed action at the $\l$-order.

\section{ Deformation of supersymmetry due to UV divergences} \label{sec:6}
\subsection{Preserving E7}
Consistency of supergravity  with unbroken local $\cH$-symmetry requires that E7 and supersymmetry  commute modulo equations of motion.  This is a consequence of the requirement that the classical or deformed action is invariant under local supersymmetry off shell $\delta _S S=0$ and under E7 on shell $\delta_{E7} S |_{S_{,i}=0}= 0$,
\be
[\delta_{E7}, \delta _S] |_{S_{,i}=0}= 0.
\label{algebra}\ee
If there is a UV divergence we add a CT to the classical action. Classical  supersymmetry transformation of the fermions depends on E7 invariant $\cH$-symmetry covariant graviphoton $\cF^{\rm (cl)}$. 
\be
 \delta \chi_{\a ijk}^{\rm cl}=  \cF_{\a\b [ij} ^{(\rm cl)}\epsilon^\b_{k]} +\dots,
\label{cl}\ee
where $\dots$ are vector-independent  dependent terms. But the classical graviphoton $\cF^{(cl)}$ is not E7 invariant  anymore. It is the deformed graviphoton
\be
\cF^{\rm def}= \cF^{\rm(cl)} + \l \hat \cF.
\ee
 which we defined in eqs. \rf{dg}, \rf{calM},   which is E7 invariant.

If we would like to preserve the E7 invariance of the fermions after supersymmetry transformations, we need to deform the supersymmetry transformation of the fermions due to UV divergence.
\be
\delta^{\rm def} \chi_{\a ijk} = \delta^{\rm cl} \chi_{\a ijk} + \hat \delta \chi_{\a ijk}
\label{deltachi}\ee
where
\be
 \hat \delta \chi_{\a ijk}= \l \hat \cF_{\a\b [ij} \epsilon^\b_{k]},
\ee
We use the  $\l$-order CT in \rf{pertCT} in spinorial notation
\begin{equation}
\cL ^{\rm CT}= {\cal F}^{ \a \b }_{  ij }  X_{\a \b \dot \a \dot \b}  \bar  {\cal F}^{\dot \a \dot \b ij } 
\label{CTsp}\end{equation}
and we find that 
\be
\hat \cF_{a\b ij}=-\cM_{ij,mn}  X_{\a \b \dot \a \dot \b} \bar \cF^{\dot \a \dot \b mn \, cl}+{\cal O}(\lambda).
\ee

Classical action is invariant under classical supersymmetry transformations, however, when we deformed classical supersymmetry transformations for spin 1/2  fermions,  we have an extra term in the supersymmetry transformation of the action
\be
\hat \delta  S^{\rm (cl)} = {\delta S^{\rm (cl)} \over \delta \chi_{ \a ijk} }\l \hat \cF_{\a\b [ij} \epsilon^\b_{k]}   +{\rm h.c.}\ .
\label{def}\ee
How to cancel this term?  There are two possibilities, the first one  is to deform some supersymmetries of the classical fields in the classical action.
The second one is
to find out if the $\l$ order CT has an analogous term to cancel \rf{def}. 

For our purpose here it is convenient to use the form of supersymmetry transformations in  \cite{Cremmer:1979up} which is manifestly \E\ covariant. In particular the supersymmetry of vectors is presented in the form of the doublet using all 56 vectors, 28 vectors $B_\mu ^{MN}$ and 28 $C_{\mu MN}$
in notations in  \cite{Cremmer:1979up} as shown in eq. (8.23) there. By checking all supersymmetry rules in eqs. (8.21)-(8.25) we can see that only fermion rules in presence of CT deformation break \E. All  supersymmetry transformations of bosons do not change due to presence of the CT in the action. Even the ones for vectors, due to a manifest doublet form of supersymmetry rules in  eq. (8.23) in  \cite{Cremmer:1979up}, have a build-in dependence on presence of the CT.

The term we would like to cancel  in eq. \rf{def} is
\be
\hat \delta_S  S^{\rm (cl)} = {\delta S^{\rm (cl)} \over \delta \chi_{ \a ijk} }\l \cM_{ij,mn}  X_{\a \b \dot \a \dot \b} \bar \cF^{\dot \a \dot \b mn} \,  \epsilon_k^\beta
   +{\rm h.c.}\ .
\label{def1}\ee
All expressions in \rf{def1} are \E\ invariant with exception of  $
\cM_{ij,kl}$. This expression,    $\cM_{ij,kl}= (u_{ij}{}^{IJ}+v_{ijIJ})(u^{kl}{}_{IJ}+v^{klIJ})^{-1}$,  is $\cH$-symmetry covariant but not \E\ invariant. Note that in both factors in $\cM_{ij,kl}$  we add terms which transform differently under E7 with indices $I,J$ both up and down.

This means that trying to cancel the term \rf{def1} we need to deform some of the 
supersymmetry transformations in classical action for the graviton, vectors or scalars by forcing some additional $\l$ corrections which are not \E\ invariant. We have not found any such transformations which would remove the problematic terms  in \rf{def1}.
 
 Furthermore, it would mean that we have to add a term in \rf{comp} that is not invariant under $E_{7(7)}$ in the transformation law of an $E_{7(7)}$ singlet field $\chi_{\alpha ijk}$. 
 But this would defeat the purpose of restoring  \E\  on fermions which is lost in presence of the CT. We conclude therefore that there is no consistent way to avoid supersymmetry breaking of the classical action while preserving  \E.  
 \subsection{Breaking E7}
 Here we will argue that breaking E7 leads to the same consequences as deforming supersymmetry on fermions with preservation of supersymmetry.  
 So, we keep classical supersymmetry transformations on fermions as in eq. \rf{cl} and the classical action is invariant under local supersymmetry, but E7 is broken.
 
 There are few aspects in this analysis.  We study supersymmetry algebra with account of nonlinear terms in supersymmetry transformations. We take into account the fact that local $\cH$ symmetry and rigid one in the unitary gauge differ by an E7 transformations. It means, the object  which was covariant under  local $\cH$ symmetry might break a rigid $\cH$ symmetry in the unitary gauge. We give a relevant  example of this phenomena.

1.  {\it Supersymmetry algebra} 

\noindent Breaking E7 when using a classical supersymmetry transformations without deformation also leads to a breaking of a nonlinear supersymmetry. We explain it here,  this effect can be seen via the local supersymmetry algebra.
 
If we keep classical supersymmetry and breaking E7,  this would mean that the fermion supersymmetry transformation is not affected by the CT, but inconsistencies will appear in the nonlinear supersymmetry algebra.  The supersymmetry algebra on 
 fermions at the linear level is 
 \be
\{\delta_{ 1}, \delta _{ 2} \} = \delta _{\rm Diff} + \delta _{SO_ {(3,1) }} +  \delta _{U(1) } + \cO(\chi^2)
\label{susy1}\ee
(see eq. (3.14) in \cite{Hillmann:2009zf}).
 But at the non-linear level (see eq. (3.22) in \cite{Hillmann:2009zf})
  \be
\{\delta_{ 1 s}, \delta _{ 2 s} \} =\delta_{3 s}+ \delta _{\rm Diff} + \delta _{SO_ {(3,1) }} +  \delta _{U(1) } +\delta _{SU(8)} 
\label{susy2}\ee
one can see that there is, in addition,  another supersymmetry transformation $ \delta _{S3}$  as well as an $SU(8)$ rotation. We have shown this $SU(8)$ rotation before in eqs. \rf{al}, \rf{curv}. Thus
 in the unitary gauge the commutator of two non-linear classical supersymmetry variations generates another supersymmetry,  a field dependent $SU(8)$ symmetry in addition to the parts which were seen in the linear approximation
 \be
\{\delta_{1s}, \delta _{2s} \} = \delta _{3s} + \delta_{SU(8)} +\dots
\label{susy3}\ee
where $\dots $ is for terms in the algebra one can see in the linear approximation.
In the unitary gauge, the rigid field dependent $SU(8)$ symmetry is a mix of locally gauge-fixed $SU(8)$ and the $SU(8)$ subgroup of E7.
Therefore, if E7 was broken before gauge-fixing, in the unitary gauge the field dependent $SU(8)$ will be broken. This results in breaking of nonlinear supersymmetry according to the algebra in eq. \rf{susy2}.

2. {\it Compensating $\cH$ symmetry transformation preserving the unitary gauge}

\noindent When we make an \E\  transformation for example in $\cN=8$, it will by itself not keep  70 scalars intact, its role is to mix 133 of them, so E7 by itself will break the unitary gauge condition that $\cV=\cV^{\dagger}$. Therefore the important feature of $SU(8)$ symmetry in the unitary gauge, is  that it  is the rigid  subgroup of $E_{7(7)} \times SU(8)$. Comparing it with the one before gauge-fixing   one finds that it involves an additional  compensating $SU(8)$ rigid field dependent transformation preserving the unitary gauge. The explicit form of this compensating transformation on fermions as a function of \E\  parameters is presented in eqs. (4.31)-(4.34) in  \cite{Kallosh:2008ic}. 

This explains why an expression  which was was covariant under  local $\cH$ symmetry might break a rigid $\cH$ symmetry in the unitary gauge.
 
3. {\it Example}

\noindent To exemplify this statement consider the expression  causing supersymmetry breaking in eq.  \rf{def1} due to
\be
\hat \delta  \chi_{\a ijk}  = \l \cM_{[ij,mn}  X_{\a \b \dot \a \dot \b} \bar \cF^{\dot \a \dot \b mn } \epsilon_{k]}^\beta
\label{def2}\ee 
When local $\cH$-symmetry is not gauge-fixed, the rhs of eq. \rf{def2} is $SU(8)$ covariant since  each factor in the product 
\be
{\cal M}_{ij,kl}\equiv (u_{ij}{}^{IJ}+v_{ijIJ})(u^{kl}{}_{IJ}+v^{klIJ})^{-1} = (u+v) (\bar u +\bar v)^{-1}
\label{calM1}\ee 
 is  $SU(8)$ covariant, although not E7 invariant, according to eqs. \rf{gauge}, \rf{Vtransf}.
 The sub-matrices $u$ and $v$ carry indices of both \E\ and $SU(8)$ ($I, J = 1,...,8, \, \, i, j = 1,...,8$) but in the unitary gauge where 
\be
\cV=\cV^{\dagger}
\ee
 we retain only manifest invariance with respect to the rigid diagonal subgroup of $E_{7(7)} \times SU(8)$, without distinction between the two types of indices.

 In the unitary gauge  \cite{Cremmer:1979up,deWit:1982bul,Kallosh:2008ic}    
  \begin{eqnarray}\label{cVu}
{\cal V}=\left(
                                        \begin{array}{cc}
                                          u_{ij} {} ^{IJ}& v_{ijKL} \\
                                           v^{klIJ} &  u^{kl}{}_{KL} \\
                                        \end{array}
                                      \right)|_{\cV=\cV^{\dagger} }\, \,    \Rightarrow \left(
                                        \begin{array}{cc}
                                         P^{-1/2} & \, \,  - (P^{-1/2} ) y\\
                                           - \bar P^{-1/2}  \bar y &   \bar P^{-1/2}  \\
                                        \end{array}
                                      \right) +\dots                                                                           
                                       \end{eqnarray}
where 
\be
P= 1-y\bar y\, , \qquad y_{ij, kl}= \phi_{ijmn}\left ( {\tanh \sqrt{{1\over 8}\bar \phi  \phi}\over \sqrt{\bar \phi \phi}}\right) ^{mn}{}_{kl}\, ,\ee                                       
Here $\phi_{ijkl}$ and $\bar \phi^{ijkl}=\pm {1\over 24} \epsilon^{ijklmnpq} \phi_{mnpq}$ transform in 35-dimensional representation of $SU(8)$. These are 70 physical scalars in the unitary gauge.

In the linear approximation $P=\bar P=1$, $y={1\over \sqrt 8} \phi$, and 
we find  
\be
u+v\Rightarrow   1-{1\over \sqrt 8} \phi\, , \qquad      \bar u+ \bar v\Rightarrow   1- {1\over \sqrt 8} \bar \phi
\ee
\be
{\cal M} |_{ \cV=\cV^{\dagger} }\, \,    \Rightarrow   I - {1\over \sqrt 8}(\phi - \bar \phi ) +\cdots
\label{calMl}\ee 
If we use indices, we see that $SU(8)$ is broken down to $SO(8)$                                    
\be
{\cal M}_{ij,kl} |_{ \cV=\cV^{\dagger} }\, \,    \Rightarrow   \delta_{ijkl} - {1\over \sqrt 8}(\phi_{ijkl} - \bar \phi ^{klij}) +\cdots
\label{calMl}\ee 

We see here that in the unitary gauge ${\cal M}_{ij,kl}$ is not $SU(8)$ covariant anymore since the $SU(8)$ symmetry in the unitary gauge (the rigid diagonal subgroup of $E_{7(7)} \times SU(8)$) has inherited the broken E7  symmetry of ${\cal M}_{ij,kl}$ before gauge-fixing.

One can also try to invent some deformation of supersymmetry/duality  rules to see how exactly to make the action invariant under local supersymmetry off shell and under \E\ on shell so that eq. \rf{al} is valid for a deformed theory. For example, we could add terms that have more powers of fields. However, addition of such terms does not alter our conclusion since it does not cancel the problematic term discussed above.

\subsection{Supersymmetry transformation of the CT} 

The CT action under classical supersymmetry is supersymmetric if classical equations of motion are satisfied. Therefore it is hard to see how the expression in \rf{def} can be compensated if the action is $S^{\rm cl}+ \l S^{\rm CT}$.

One can look at a simple example when we keep only the terms in variation \rf{def} with the minimal number of fields, both in fermion eq. of motion as well as in $\hat \cF$
\be
\hat \delta  S^{\rm (cl)} = \l ( \partial^{\a}{}_{ \dot \a} \bar \chi^{ \dot \a ijk} )\delta _{im} \delta _{jn} X_{\a \b \dot \a \dot \b} \bar \cF^{\dot \a \dot \b mn \, cl} \epsilon^\b_{k} +\dots ,
\label{linear}\ee
where we have used the fact that $M_{ij,mn}=\delta_{im}\delta_{jn}+\cdots$ at the leading order in scalar fields expansion. If we choose $X_{\a \b \dot \a \dot \b}$ depending on 2 gravitons, this is a 4-field expression with one fermion, 2 graviton and one vector.\footnote{We have used $SU(8)$ covariant operator $X_{ij}{}^{kl}$ but now we are focusing on the terms containing two gravitons with two fermions or with two vectors, and the $SU(8)$ connection does not contribute. Therefore, $X$ becomes an operator without $SU(8)$ indices independently of our choice of the full $X_{ij}{}^{kl}$.} Then, the expansion of $X$ in fields is given by
\be
X_{ \a \b \, \dot \a \dot \b }= R_{\a \b \gamma \delta}(x) R_{\dot \a \dot \b \dot \gamma \dot \delta} (x)\partial ^{2(L-3)} \partial ^{\gamma \, \dot \gamma } \partial ^{ \delta \, \dot \delta }+\cdots \, ,
\ee
where ellipses denote the operators that have more fields. Using the leading order expansion, the CT we consider is reduced as
\begin{equation}
\cL ^{CT}= {\cal F}^{ \a \b }_{  ij }  X_{\a \b \dot \a \dot \b}  \bar  {\cal F}^{\dot \a \dot \b ij } ={\cal F}^{ \a \b }_{  ij } R_{\a \b \gamma \delta}(x) R_{\dot \a \dot \b \dot \gamma \dot \delta} (x)\partial ^{2(L-3)} \partial ^{\gamma \, \dot \gamma } \partial ^{ \delta \, \dot \delta }\bar  {\cal F}^{\dot \a \dot \b ij }+\cdots
\end{equation}
where the ellipses denote terms having more fields. This leading term corresponds to a linearized supersymmetric $R^4$ invariant in \cite{Freedman:2011uc} with extra $\partial^{2(L-3)}$ derivatives. We would like to ask whether the linearized  supersymmetric $\partial^{2(L-3)}R^4$ has a term that cancel the supersymmetry variation~\eqref{linear} since it has the same numbers of fields. Indeed, there is the 2-fermion, 2 graviton term in the linearized superinvariant in \cite{Freedman:2011uc} where (in line 5 in eq. (6.8))
\be
R_{\dot \a \dot \b \dot \gamma \dot \delta} \bar \chi^{\dot \a ijk} \partial^{\b \dot \b} \partial^{\gamma \dot \gamma}  \partial^{\delta \dot \delta} \partial^{2(L-3)} \chi^{\a}_{ ijk} R_{\a \b \gamma \delta}.
\ee
However, the supersymmetry transformation of this term does not have the tensorial structure to cancel~\eqref{linear}, and therefore, the additional supersymmetry variation due to  deformation of the graviphoton cannot be canceled.

Let us repeat and summarize our logic here: First, we have considered a deformation of twisted duality condition so that we keep $E_{7(7)}$ invariance. This leads to a deformation of the graviphoton, and accordingly the deformation of fermion's supersymmetry variation. Thus, we find an additional supersymmetry variation~\eqref{linear}. On the other hand, the CT we have added at ${\cal O}(\lambda)$ can be represented by  a linearized supersymmetric CT shown in \cite{Freedman:2011uc}. We have asked whether the supersymmetry transformation of the linearized superinvariant can cancel the additional variation~\eqref{linear} associated with the graviphoton variation. We have found  that it does not cancel the supersymmetry breaking of thew classical action due nto deformation of the supersymmetry on fermions.

The deformation of local supersymmetry transformation is required to preserve the E7 invariance of the fermions after supersymmetry transformation. Thus if it is not cancelled within the action $S^{\rm cl}+ \l S^{\rm CT}$, it means that this action, which is claimed to restore E7 symmetry  in presence of fermions,  is not invariant under deformed supersymmetry.

If we do not deform the supersymmetric transformation of the fermions, the problem we referred to does not arise directly.  However, it does arise as we have shown using the supersymmetry algebra and the fact that the effect of broken E7 before gauge-fixing $\cH$-symmetry leads to a broken rigid $\cH$-symmetry and breaking of non-linear local supersymmetry via the algebra in eq. \rf{susy2}.

It appears we have a choice, if UV divergence will be detected: save either supersymmetry of the deformed action, or E7 symmetry. But not both of them. Moreover, the broken E7 symmetry with preserved local $\cH$-symmetry leads to a broken local nonlinear supersymmetry in the unitary gauge. Either way, UV divergence leads to breaking of local nonlinear supersymmetry.

\noindent Finally, we would like to add some  comments here:
\begin{itemize}
\item We have focused mostly  on  UV divergence at 4-point interactions, particularly by the reason that 4-point loop computations of UV divergences might be expected. However,  there are candidate CT's   at higher point interactions, independent on 4-point UV divergences.  These may be studied in the future.

\item We would like to emphasize here that we made a  choice of the 
 deformed twisted duality constraint \rf{defTDC} associated with the candidate CT ~\eqref{pertCT} at the leading order. If we would try to make a more general choice  of the 
 deformed twisted duality constraint associated with the candidate CT ~\eqref{pertCT} it might affect the form of the all order in $\l$
 deformed action. However, it would not affect the analysis of local supersymmetry breaking at the order $\l$. Therefore the conclusion about UV divergence leading to local nonlinear supersymmetry breaking  is not affected by our choice of the deformed twisted duality constraint. 
\end{itemize}

\section{ Discussion and Summary}\label{sec:7}
 $\N\geq 5$ supergravities have local $\cH$ symmetry and global E7-type on shell symmetry, in addition to nonlinear local supersymmetry, see for example $\cN=8$ case in \cite{Cremmer:1979up,deWit:1982bul} with 133 scalars, 70 of which are physical.  
When these supergravities are described in a form when local $\cH$ symmetry is not gauge-fixed (like in $\cN=8$ with 133 scalars), both  local $\cH$ symmetry and global E7-type on shell symmetry are independent and linearly realized. The nonlinear local supersymmetries are  E7  and  $\cH$-symmetry covariant, see for example eqs. (8.21)-(8.25) in \cite{Cremmer:1979up}. Moreover, local supersymmetry and E7 symmetry commute, modulo equations of motion.
 
 In the unitary gauge  where all  parameters of local $\cH$ symmetry are used to eliminate unphysical scalars (63 in $\cN=8$), global E7 symmetry is nonlinearly realized and the remaining rigid $\cH$ symmetry is a mix of originally independent  local $\cH$ symmetry and global E7-type on shell symmetry.

In our goal to study deformation of  $\N\geq 5$ supergravities with the purpose to absorb potential UV divergences it was important to generalize the results in \cite{Kallosh:2018mlw} where the deformation of duality was described without fermions,  using symplectic formalism.   This is effectively a unitary gauge where the local $\cH$ symmetry is gauge-fixed and there are only physical scalars (70 in $\cN=8$).

We have introduced a deformed  twisted   self-duality constraint in \rf{defTDC}. In absence of fermions and in the unitary gauge with only physical scalars our constraint is reduced to the one in eq. (3.4) of \cite{Kallosh:2018mlw}. Here we have constructed the action including all order in $\l$ corrections  with on shell deformed duality symmetry, local  $\cH$ symmetry and with fermions present. In particular, the solution of the $\l$-corrected twisted   self-duality constraint gives an expression for the all order in $\l$ deformed   graviphoton in eq. \rf{allorder}. It is covariant under local $\cH$-symmetry, it is  E7 duality invariant, if all order in $\l$ are taken into account. 

This means that if we only take the first order correction in $\l$ to deform the action, E7 symmetry of the theory is broken starting  at the level  $\l^2$, as it was shown in \cite{Kallosh:2011dp} and confirmed in \cite{Bossard:2011ij,Kallosh:2018mlw}.  Therefore it was possible to add to the action higher order in $\l$ terms, starting with $\l^2$ terms and restore the deformed duality symmetry. We now have seen here that the same is possible before gauge-fixing $\cH$ symmetry and with fermions present.

In presence of fermions with unbroken local $\cH$-symmetry we were able to ask the question about the local supersymmetry of the deformed action at the order $\l$.  The E7 invariant  $\cH$-covariant fermions, under supersymmetry transform into E7 invariant $\cH$-covariant graviphoton, classically. Once the deformation was added  to   absorb  the  UV divergence, the graviphoton is deformed to restore  E7 invariance. This deformation affects the fermions supersymmetry transformations, see eqs. \rf{deltachi}-\rf{def}. We explained why the deformed action breaks local supersymmetry at the level $\l$ and why there is no way to restore it, as opposite to E7 symmetry which was unbroken at  the level $\l$. If we make a choice to break instead E7 symmetry by not deforming supersymmetry transformations of the fermions, we find that it leads to broken supersymmetry anyway, because    the  rigid $\cH$-symmetry is a mix of a local $\cH$-symmetry and E7 symmetry.

\noindent {\it To summarize, our results on $\N\geq 5$,  $d=4$ supergravities are the following}.

We have recalled  the fact  that the CT's proposed in  \cite{Kallosh:1980fi} are of two types: the linearized at $L\leq \cN-1 $ which cannot be promoted to a level where they have nonlinear local supersymmetry,   and the ones, at $L\geq  N $ which have an on shell nonlinear supersymmetry.

1. We have explained the enhanced ultraviolet cancellation of 82 diagrams in UV divergence in $\cN=5, L=4$ in  \cite{Bern:2014sna} using local nonlinear supersymmetry. We have  pointed out that the CT proposed in \cite{Bossard:2011tq} in harmonic superspace breaks local nonlinear supersymmetry, despite it has linearized supersymmetry, see eq. \rf{N5} and discussion around this formula.

2. We have explained (in Appendix \ref{appA}, since this work is about $\cN\geq 5$)  the enhanced ultraviolet cancellation of UV divergence in $\cN=4, L=3$ in  \cite{Bern:2012cd} using local nonlinear supersymmetry. The 1-loop $U(1)$ anomaly \cite{Carrasco:2013ypa} of this theory is also a local nonlinear supersymmetry anomaly, as well as a local superconformal anomaly. This explains the structure of  the  UV divergence at $L=4$  in \cite{Bern:2013uka}.
The 3d case of enhanced cancellation in $d=5$ is explained also via nonlinear supersymmetry in \cite{RK}.

3.  If UV divergences will show up at $L < L_{\rm cr}= \cN$ ($\cN=6, L=5$ and $\cN=8, L=7$) they will also qualify as quantum corrections breaking nonlinear local supersymmetry, as we elaborated in Sec. \ref{sec:41}.  This statement follows from dimensional analysis and properties of geometric candidate CT's and the fact that these do not exist at $L < L_{\rm cr}= \cN$ \cite{Kallosh:1980fi,Howe:1980th}. The linearized ones, which exist at $L < L_{\rm cr}= \cN$ in the  unitary gauge, break nonlinear local supersymmetry.
 
4.  If UV divergences will show up at $L \geq  L_{\rm cr}= \cN$, they will also qualify as quantum corrections breaking nonlinear local supersymmetry, as we elaborated in Sec. \ref{sec:42} and in Sec. \ref{sec:6}.  The proof of this result, however, required    a more significant effort compared to $L < L_{\rm cr}= \cN$ cases. Namely we had to study deformation of E7 symmetry and deformation of local supersymmetry before gauge-fixing local $\cH$-symmetry. In such case local $\cH$ symmetry and global E7 symmetry are independent and both linearly realized.
 This was done in Secs. \ref{sec:5},  \ref{sec:6}.  We have found that the deformed action breaks deformed nonlinear supersymmetry, either directly, or indirectly via broken E7 symmetry which reflects on nonlinear local supersymmetry.  

\

In conclusion, from the loop computations available,  we know that  $\cN=5, L=4$ supergravity is UV finite    \cite{Bern:2014sna}.  Here it is now explained by the requirement of unbroken local nonlinear supersymmetry since the harmonic superspace candidate CT \cite{Bossard:2011tq} is not valid at the nonlinear level.
If more $L < L_{\rm cr}= \cN$ loop computations will be available and   will be  UV finite, for example, $\cN=6, L=5$ and $\cN=8, L=7$, the same nonlinear  local supersymmetry argument explaining UV finiteness  will work, since the harmonic superspace candidate CT \cite{Bossard:2011tq} for $L=\cN-1$ is not valid at the nonlinear level. 

However, at present there are no examples of $L \geq  L_{\rm cr}= \cN$ loop computations. If $\cN=5, L=5$ supergravity will be  found  to be UV divergent, we will conclude  that the relevant  deformed supergravity is BRST inconsistent since  nonlinear local supersymmetry of the deformed action is broken. But if $\cN=5, L=5$  will be found to be UV finite, it will be  explained by unbroken nonlinear local supersymmetry arguments in Secs.~\ref{sec:42}, \ref{sec:6}. 

This will support the earlier work  where UV finiteness was predicted based on manifest E7 symmetry \cite{Kallosh:2018wzz}, or  on  properties of the unitary conformal supermultiplets \cite{Gunaydin:2018kdz}, assuming unbroken supersymmetry. Here we have investigated nonlinear local supersymmetry directly.

\section*{Acknowledgement}
We are grateful to our collaborators on   earlier related projects:  S. Ferrara, D. Freedman, M. Gunaydin, H. Nicolai, T. Ortin, A. Van Proeyen. We had extremely useful discussions of the current work with J. J. Carrasco and  R. Roiban. It is our understanding from Z. Bern and J. J. Carrasco that the computation of the UV divergence in $\cN=L=5$ might be possible in the future, which stimulated our work.

RK   is supported by SITP and by the US National Science Foundation grant PHY-2014215.
YY is supported by Waseda University Grant for Special Research Projects (Project number: 2022C-573).
\appendix

\section{$d=4, \, \cN=4$ enhanced cancellation and nonlinear supersymmetry anomaly}\label{appA}
Since the purpose of this work is to study $\cN\geq 5$ supergravities, we have put the new developments  in $\cN=4, d=4$ in the Appendix. It is however, a reflection of what we have learned in $\cN\geq 5$ supergravities.

As discussed above in Sec. \ref{sec:42} the candidate CT's with nonlinear local supersymmetry are available starting  $L=\cN$ \cite{Kallosh:1980fi,Howe:1980th}. The harmonic space CT proposed in \cite{Bossard:2011tq} in this case has the same problems we discussed in Sec. \rf{sec:12}. Namely the proof of consistency of the harmonic superspace in \cite{Hartwell:1994rp} above the linear level is not available for $\cN=4$ Poincar\'e supergravity.

This explains why there is an enhanced cancellation  in $\cN=4, L=3$ \cite{Bern:2012cd}: the CT's with local nonlinear supersymmetry exist only starting from $L=4$ and is absent in $L=3$. The linearized CT at $L=3$ is
\be
CT^{L=3}_{\rm lin}= \kappa^4\int d^4 x \, d^{16} \theta \, (W\bar W)^2
\label{L3ctN4}\ee
The zero dimension chiral superfield $W$ and its conjugate anti-chiral superfield $\bar W$ break nonlinear supersymmetry, as we explained in Sec. \ref{sec:13}, although the superinvariant in \rf{L3ctN4} has linearized $\cN=4$ supersymmetry.

This theory has 1-loop amplitude anomalies \cite{Carrasco:2013ypa} and is UV divergent at $L=4$ \cite{Bern:2013uka}. It is interesting that at $L=3$ anomaly has not yet kicked in\footnote{We believe it is possible to  explain it  using superconformal version of this theory \cite{Ferrara:2012ui} which also sheds the light on the common irrational factor in front of all 3 UV divergences at $L=4$.}. In $L=4$ the UV divergences are given by 3 different superinvariants  \cite{Bern:2013uka} of Poincar\'e $\cN=4$ supergravity. Only one of them has full nonlinear supersymmetry, see the general case in eq. \rf{CT8}.
\be
CT^{L=4}_{ 1\, \rm nonlin}=\kappa^6 \int d^4 x \, d^{16} \theta  \det E \,   \chi_{\alpha}^i \,  \chi^{\alpha j} \,  \bar \chi_{\dot \alpha \,  i} \bar \chi^{ \dot \alpha }_j    = \kappa^{6} \int d^4 x \, D^{2} \, R^4 +\dots
\label{L4}\ee
The additional 2 UV divergences discovered in \cite{Bern:2013uka} have found to have the same structure as  $U(1)$ anomalies in \cite{Carrasco:2013ypa}. 
Namely, the 1-loop $U(1)$ anomalies in  \cite{Carrasco:2013ypa} are described by the following linearized chiral superspace invariants
\be
 {\rm Anomaly}^{L=1}_{ 2\, \rm lin} \to   \int d^4 x\, d^{8} \theta \, W^2  W^2 \pm  hc\,,
\label{I42a}\ee
\be
 {\rm Anomaly}^{L=1}_{3 \, \rm lin} \to  \int d^4 x\, d^{8} \theta \, \bar C_{\dot \alpha \dot \beta \dot \gamma \dot \delta} W \partial^{\alpha \dot \alpha} \partial^{\beta \dot \beta} W \partial^{-6} \partial_{\alpha}^{ \dot \gamma} \partial_{\beta}^{ \dot \delta} W \pm hc \,,
\label{I43a}\ee
Now we can present them here as 4-loop CT's which have linearized supersymmetry and break  nonlinear supersymmetry. Namely, uplifting the 1-loop nonlocal anomaly structures in \cite{Carrasco:2013ypa} by $ \kappa^6 stu$ we present local CT's, UV divergences in $L=4$
\be
 {\rm CT}^{L=4}_{2\, \rm lin} \to  \kappa^{6} \int d^4 x\, d^{8} \theta \, W^2 \partial^6 W^2 + {\rm h.c.}\,,
\label{I42}\ee
\be
 {\rm CT}^{L=4}_{3\, \rm lin} \to  \kappa^{6} \int d^4 x\, d^{8} \theta \, \bar C_{\dot \alpha \dot \beta \dot \gamma \dot \delta} W \partial^{\alpha \dot \alpha} \partial^{\beta \dot \beta} W \partial_{\alpha}^{ \dot \gamma} \partial_{\beta}^{ \dot \delta} W + \rm{ h.c.} \,.
\label{I43}\ee
These break nonlinear supersymmetry since there is no generalization of these two linear superinvariants to the nonlinear level. In  \cite{Carrasco:2013ypa} these 1-loop linearized superinvariants were discovered with the purpose to expose $U(1)$  anomaly. This $U(1)$ is a subgroup of the duality $SL(2, \mathbb{R})$ symmetry which was broken.  Note that  $SL(2, \mathbb{R})$  is also group of type E7 \cite{Ferrara:2011dz}.

Thus here again we see that breaking E7-type  $SL(2, \mathbb{R})$ duality means also  breaking of a local nonlinear supersymmetry. Accordingly, the 1-loop $U(1)$  anomaly is related to 1-loop local nonlinear supersymmetry anomaly. 
These two expressions in \rf{I42a}, \rf{I43a}  represent the {\it $U(1)$ anomaly as well as nonlinear supersymmetry anomaly}: they are given by subspace of the superspace superinvariants which do not have a nonlinear generalization.

We find it now extremely plausible that all these properties of   $\cN=4$  Poincar\'e supergravity come from the superconformal version of the theory, as discussed in \cite{Ferrara:2012ui}. This superconformal theory has anomaly defined by one structure combining tree independent $\cN=4$  Poincar\'e supergravity $L=4$ UV divergences. If this is the case, the reason why at $L=3$ there are no UV divergences is that in addition to the fact that there is no nonlinear candidate CT, the absence of anomaly is also explained: the CT in \rf{I43} at the 3-loop order is non-local. Therefore since all 3 UV divergences correspond to one expression in superconformal theory,  breaking of superconformal symmetry did not show up  at $L=3$ but only at $L=4$ where the CT in \rf{I43} at the 4-loop order is local.

\section{Identities for $E_{7(7)}/SU(8)$ matrices}\label{appB}
We summarize some identities for the $E_{7(7)}/SU(8)$ matrices given also in \cite{deWit:1982bul}.
\begin{align}
&u_{ij}{}^{IJ}u^{kl}{}_{IJ}-v_{ijIJ}v^{klIJ}=\delta_{ij}^{kl},\\
&u_{ij}{}^{IJ}v_{klIJ}+v_{ijIJ}u_{kl}{}^{IJ}=0,\\
&u^{ij}{}_{IJ}v^{klIJ}-v^{ijIJ}u^{kl}{}_{IJ}=0,\\
&u^{ij}{}_{IJ}u_{ij}{}^{KL}-v_{ijIJ}v^{ijKL}=\delta_{IJ}^{KL},\\
&u^{ij}{}_{IJ}v_{ijKL}-v_{ijIJ}u^{ij}{}_{KL}=0,\\
&v^{ijIJ}(u^{-1})^{KL}{}_{ij}-(u^{-1})^{IJ}{}_{ij}v^{ijKL}=0,\label{A6}\\
&u_{ij}{}^{KL}-v_{ijIJ}(u^{-1})^{KL}{}_{kl}v^{klIJ}=(u^{-1})^{KL}{}_{ij}.
\end{align}
In  \cite{deWit:1982bul}, a matrix $S^{IJ,KL}$ is introduced, which can be identified as
\begin{equation}
S^{IJ,KL}\equiv (u^{ij}{}_{IJ}+v^{ijIJ})^{-1}u^{ij}{}_{KL}.\label{Sdefinition}
\end{equation}
This matrix satisfies the following identities
\begin{align}
&(u^{ij}{}_{IJ}+v^{ijIJ})S^{IJ,KL}=u^{ij}{}_{KL},\\
&(S^{-1}-{\mathbf1})^{IJ,KL}=(u^{-1})^{IJ}{}_{ij}v^{ijKL}=(u^{-1})^{KL}{}_{ij}v^{ijIJ},
\end{align}
where $\mathbf{1}$ denotes an identity $\delta^{IJ}_{KL}$, the first one follows from the definition and the second follows from \eqref{A6}.
 
\bibliographystyle{JHEP}
\bibliography{refs}

\providecommand{\href}[2]{#2}\begingroup\raggedright\begin{thebibliography}{10}

\bibitem{Kallosh:1974yh}
R.~E. Kallosh, \emph{{The Renormalization in Nonabelian Gauge Theories}},
  \href{http://dx.doi.org/10.1016/0550-3213(74)90284-3}{\emph{Nucl. Phys.} {\bf
  B78} (1974) 293--312}.

\bibitem{Goroff:1985th}
M.~H. Goroff and A.~Sagnotti, \emph{{The Ultraviolet Behavior of Einstein
  Gravity}}, \href{http://dx.doi.org/10.1016/0550-3213(86)90193-8}{\emph{Nucl.
  Phys.} {\bf B266} (1986) 709--736}.

\bibitem{Gibbons:1978ac}
G.~W. Gibbons, S.~W. Hawking and M.~J. Perry, \emph{{Path Integrals and the
  Indefiniteness of the Gravitational Action}},
  \href{http://dx.doi.org/10.1016/0550-3213(78)90161-X}{\emph{Nucl. Phys. B}
  {\bf 138} (1978) 141--150}.

\bibitem{Abreu:2020lyk}
S.~Abreu, F.~Febres~Cordero, H.~Ita, M.~Jaquier, B.~Page, M.~S. Ruf et~al.,
  \emph{{Two-Loop Four-Graviton Scattering Amplitudes}},
  \href{http://dx.doi.org/10.1103/PhysRevLett.124.211601}{\emph{Phys. Rev.
  Lett.} {\bf 124} (2020) 211601}, [\href{http://arxiv.org/abs/2002.12374}{{\tt
  2002.12374}}].

\bibitem{Becchi:1975nq}
C.~Becchi, A.~Rouet and R.~Stora, \emph{{Renormalization of Gauge Theories, \,
  CPT-1975-P.723}},
  \href{http://dx.doi.org/10.1016/0003-4916(76)90156-1}{\emph{Annals Phys.}
  {\bf 98} (1976) 287--321}.

\bibitem{Tyutin:1975qk}
I.~V. Tyutin, \emph{{Gauge Invariance in Field Theory and Statistical Physics
  in Operator Formalism, \, LEBEDEV-1975-39}},
  \href{http://arxiv.org/abs/0812.0580}{{\tt 0812.0580}}.

\bibitem{Cremmer:1979up}
E.~Cremmer and B.~Julia, \emph{{The SO(8) Supergravity}},
  \href{http://dx.doi.org/10.1016/0550-3213(79)90331-6}{\emph{Nucl. Phys.} {\bf
  B159} (1979) 141--212}.

\bibitem{deWit:1982bul}
B.~de~Wit and H.~Nicolai, \emph{{N=8 Supergravity}},
  \href{http://dx.doi.org/10.1016/0550-3213(82)90120-1}{\emph{Nucl. Phys.} {\bf
  B208} (1982) 323}.

\bibitem{Kallosh:1980fi}
R.~E. Kallosh, \emph{{Counterterms in extended supergravities}},
  \href{http://dx.doi.org/10.1016/0370-2693(81)90964-3}{\emph{Phys. Lett.} {\bf
  B99} (1981) 122--127}.

\bibitem{Howe:1980th}
P.~S. Howe and U.~Lindstrom, \emph{{Higher Order Invariants in Extended
  Supergravity}},
  \href{http://dx.doi.org/10.1016/0550-3213(81)90537-X}{\emph{Nucl. Phys.} {\bf
  B181} (1981) 487--501}.

\bibitem{Brink:1979nt}
L.~Brink and P.~S. Howe, \emph{{The $N=8$ Supergravity in Superspace}},
  \href{http://dx.doi.org/10.1016/0370-2693(79)90464-7}{\emph{Phys. Lett.} {\bf
  88B} (1979) 268--272}.

\bibitem{Howe:1981gz}
P.~S. Howe, \emph{{Supergravity in Superspace}},
  \href{http://dx.doi.org/10.1016/0550-3213(82)90349-2}{\emph{Nucl. Phys.} {\bf
  B199} (1982) 309--364}.

\bibitem{Brink:2008qc}
L.~Brink, S.-S. Kim and P.~Ramond, \emph{{$E_{7(7)}$ on the Light Cone}},
  \href{http://dx.doi.org/10.1088/1126-6708/2008/06/034}{\emph{JHEP} {\bf 06}
  (2008) 034}, [\href{http://arxiv.org/abs/0801.2993}{{\tt 0801.2993}}].

\bibitem{Kallosh:2009db}
R.~Kallosh, \emph{{N=8 Supergravity on the Light Cone}},
  \href{http://dx.doi.org/10.1103/PhysRevD.80.105022}{\emph{Phys. Rev.} {\bf
  D80} (2009) 105022}, [\href{http://arxiv.org/abs/0903.4630}{{\tt
  0903.4630}}].

\bibitem{Kallosh:2010kk}
R.~Kallosh, \emph{{The Ultraviolet Finiteness of N=8 Supergravity}},
  \href{http://dx.doi.org/10.1007/JHEP12(2010)009}{\emph{JHEP} {\bf 12} (2010)
  009}, [\href{http://arxiv.org/abs/1009.1135}{{\tt 1009.1135}}].

\bibitem{Kallosh:2012yy}
R.~Kallosh and T.~Ortin, \emph{{New $E_{7(7)}$ invariants and amplitudes}},
  \href{http://dx.doi.org/10.1007/JHEP09(2012)137}{\emph{JHEP} {\bf 09} (2012)
  137}, [\href{http://arxiv.org/abs/1205.4437}{{\tt 1205.4437}}].

\bibitem{Gunaydin:2013pma}
M.~Gunaydin and R.~Kallosh, \emph{{Obstruction to $E_{7(7)}$ Deformation in N=8
  Supergravity}},  \href{http://arxiv.org/abs/1303.3540}{{\tt 1303.3540}}.

\bibitem{Kallosh:2018wzz}
R.~Kallosh, \emph{{The Action with Manifest E7 Type Symmetry}},
  \href{http://dx.doi.org/10.1007/JHEP05(2019)109}{\emph{JHEP} {\bf 05} (2019)
  109}, [\href{http://arxiv.org/abs/1812.08087}{{\tt 1812.08087}}].

\bibitem{Gunaydin:2018kdz}
M.~Gunaydin and R.~Kallosh, \emph{{Supersymmetry constraints on U-duality
  invariant deformations of $N \geq 5$ Supergravity}},
  \href{http://dx.doi.org/10.1007/JHEP09(2019)105}{\emph{JHEP} {\bf 09} (2019)
  105}, [\href{http://arxiv.org/abs/1812.08758}{{\tt 1812.08758}}].

\bibitem{Kallosh:2008ic}
R.~Kallosh and M.~Soroush, \emph{{Explicit Action of E7 on N=8 Supergravity
  Fields}},
  \href{http://dx.doi.org/10.1016/j.nuclphysb.2008.04.006}{\emph{Nucl. Phys. B}
  {\bf 801} (2008) 25--44}, [\href{http://arxiv.org/abs/0802.4106}{{\tt
  0802.4106}}].

\bibitem{Bern:2023zkg}
Z.~Bern, J.~J.~M. Carrasco, M.~Chiodaroli, H.~Johansson and R.~Roiban,
  \emph{{Supergravity amplitudes, the double copy and ultraviolet behavior}},
  \href{http://arxiv.org/abs/2304.07392}{{\tt 2304.07392}}.

\bibitem{RK}
R.~Kallosh, \emph{{Is d=4 maximal supergravity special?}},
  \href{http://arxiv.org/abs/in preparation}{{\tt in preparation}}.

\bibitem{Bern:2014sna}
Z.~Bern, S.~Davies and T.~Dennen, \emph{{Enhanced ultraviolet cancellations in
  $\mathcal N=5$ supergravity at four loops}},
  \href{http://dx.doi.org/10.1103/PhysRevD.90.105011}{\emph{Phys. Rev.} {\bf
  D90} (2014) 105011}, [\href{http://arxiv.org/abs/1409.3089}{{\tt
  1409.3089}}].

\bibitem{Bossard:2011tq}
G.~Bossard, P.~S. Howe, K.~S. Stelle and P.~Vanhove, \emph{{The vanishing
  volume of D=4 superspace}},
  \href{http://dx.doi.org/10.1088/0264-9381/28/21/215005}{\emph{Class. Quant.
  Grav.} {\bf 28} (2011) 215005}, [\href{http://arxiv.org/abs/1105.6087}{{\tt
  1105.6087}}].

\bibitem{Hartwell:1994rp}
G.~G. Hartwell and P.~S. Howe, \emph{{(N, p, q) harmonic superspace}},
  \href{http://dx.doi.org/10.1142/S0217751X95001820}{\emph{Int. J. Mod. Phys.
  A} {\bf 10} (1995) 3901--3920},
  [\href{http://arxiv.org/abs/hep-th/9412147}{{\tt hep-th/9412147}}].

\bibitem{Ferrara:2012ui}
S.~Ferrara, R.~Kallosh and A.~Van~Proeyen, \emph{{Conjecture on hidden
  superconformal symmetry of $N=4$ Supergravity}},
  \href{http://dx.doi.org/10.1103/PhysRevD.87.025004}{\emph{Phys. Rev.} {\bf
  D87} (2013) 025004}, [\href{http://arxiv.org/abs/1209.0418}{{\tt
  1209.0418}}].

\bibitem{Drummond:2008vq}
J.~M. Drummond, J.~Henn, G.~P. Korchemsky and E.~Sokatchev, \emph{{Dual
  superconformal symmetry of scattering amplitudes in N=4 super-Yang-Mills
  theory}},
  \href{http://dx.doi.org/10.1016/j.nuclphysb.2009.11.022}{\emph{Nucl. Phys. B}
  {\bf 828} (2010) 317--374}, [\href{http://arxiv.org/abs/0807.1095}{{\tt
  0807.1095}}].

\bibitem{Elvang:2013cua}
H.~Elvang and Y.-t. Huang, \emph{{Scattering Amplitudes}},
  \href{http://arxiv.org/abs/1308.1697}{{\tt 1308.1697}}.

\bibitem{Freedman:2017zgq}
D.~Z. Freedman, R.~Kallosh, D.~Murli, A.~Van~Proeyen and Y.~Yamada,
  \emph{{Absence of U(1) Anomalous Superamplitudes in $\mathcal{N}\geq 5$
  Supergravities}},
  \href{http://dx.doi.org/10.1007/JHEP05(2017)067}{\emph{JHEP} {\bf 05} (2017)
  067}, [\href{http://arxiv.org/abs/1703.03879}{{\tt 1703.03879}}].

\bibitem{Elvang:2009wd}
H.~Elvang, D.~Z. Freedman and M.~Kiermaier, \emph{{Solution to the Ward
  Identities for Superamplitudes}},
  \href{http://dx.doi.org/10.1007/JHEP10(2010)103}{\emph{JHEP} {\bf 10} (2010)
  103}, [\href{http://arxiv.org/abs/0911.3169}{{\tt 0911.3169}}].

\bibitem{Ferrara:2011dz}
S.~Ferrara and R.~Kallosh, \emph{{Creation of Matter in the Universe and Groups
  of Type E7}}, \href{http://dx.doi.org/10.1007/JHEP12(2011)096}{\emph{JHEP}
  {\bf 12} (2011) 096}, [\href{http://arxiv.org/abs/1110.4048}{{\tt
  1110.4048}}].

\bibitem{Cremmer:1998px}
E.~Cremmer, B.~Julia, H.~Lu and C.~N. Pope, \emph{{Dualization of dualities. 2.
  Twisted self-duality of doubled fields, and superdualities}},
  \href{http://dx.doi.org/10.1016/S0550-3213(98)00552-5}{\emph{Nucl. Phys. B}
  {\bf 535} (1998) 242--292}, [\href{http://arxiv.org/abs/hep-th/9806106}{{\tt
  hep-th/9806106}}].

\bibitem{Gaillard:1981rj}
M.~K. Gaillard and B.~Zumino, \emph{{Duality Rotations for Interacting
  Fields}}, \href{http://dx.doi.org/10.1016/0550-3213(81)90527-7}{\emph{Nucl.
  Phys.} {\bf B193} (1981) 221--244}.

\bibitem{Andrianopoli:1996ve}
L.~Andrianopoli, R.~D'Auria and S.~Ferrara, \emph{{U duality and central
  charges in various dimensions revisited}},
  \href{http://dx.doi.org/10.1142/S0217751X98000196}{\emph{Int. J. Mod. Phys.}
  {\bf A13} (1998) 431--490}, [\href{http://arxiv.org/abs/hep-th/9612105}{{\tt
  hep-th/9612105}}].

\bibitem{Hillmann:2009zf}
C.~Hillmann, \emph{{$E_{7(7)}$ invariant Lagrangian of d=4 N=8 supergravity}},
  \href{http://dx.doi.org/10.1007/JHEP04(2010)010}{\emph{JHEP} {\bf 04} (2010)
  010}, [\href{http://arxiv.org/abs/0911.5225}{{\tt 0911.5225}}].

\bibitem{Bossard:2010dq}
G.~Bossard, C.~Hillmann and H.~Nicolai, \emph{{E7(7) symmetry in perturbatively
  quantised N=8 supergravity}},
  \href{http://dx.doi.org/10.1007/JHEP12(2010)052}{\emph{JHEP} {\bf 12} (2010)
  052}, [\href{http://arxiv.org/abs/1007.5472}{{\tt 1007.5472}}].

\bibitem{Kallosh:2011dp}
R.~Kallosh, \emph{{$E_{7(7)}$ Symmetry and Finiteness of N=8 Supergravity}},
  \href{http://dx.doi.org/10.1007/JHEP03(2012)083}{\emph{JHEP} {\bf 03} (2012)
  083}, [\href{http://arxiv.org/abs/1103.4115}{{\tt 1103.4115}}].

\bibitem{Kallosh:2011qt}
R.~Kallosh, \emph{{N=8 Counterterms and $E_{7(7)}$ Current Conservation}},
  \href{http://dx.doi.org/10.1007/JHEP06(2011)073}{\emph{JHEP} {\bf 06} (2011)
  073}, [\href{http://arxiv.org/abs/1104.5480}{{\tt 1104.5480}}].

\bibitem{Bossard:2011ij}
G.~Bossard and H.~Nicolai, \emph{{Counterterms vs. Dualities}},
  \href{http://dx.doi.org/10.1007/JHEP08(2011)074}{\emph{JHEP} {\bf 08} (2011)
  074}, [\href{http://arxiv.org/abs/1105.1273}{{\tt 1105.1273}}].

\bibitem{Carrasco:2011jv}
J.~J.~M. Carrasco, R.~Kallosh and R.~Roiban, \emph{{Covariant procedures for
  perturbative non-linear deformation of duality-invariant theories}},
  \href{http://dx.doi.org/10.1103/PhysRevD.85.025007}{\emph{Phys. Rev.} {\bf
  D85} (2012) 025007}, [\href{http://arxiv.org/abs/1108.4390}{{\tt
  1108.4390}}].

\bibitem{Pasti:2012wv}
P.~Pasti, D.~Sorokin and M.~Tonin, \emph{{Covariant actions for models with
  non-linear twisted self-duality}},
  \href{http://dx.doi.org/10.1103/PhysRevD.86.045013}{\emph{Phys. Rev. D} {\bf
  86} (2012) 045013}, [\href{http://arxiv.org/abs/1205.4243}{{\tt 1205.4243}}].

\bibitem{Kallosh:2018mlw}
R.~Kallosh, H.~Nicolai, R.~Roiban and Y.~Yamada, \emph{{On quantum
  compatibility of counterterm deformations and duality symmetries in $
  \mathcal{N}\ge 5 $ supergravities}},
  \href{http://dx.doi.org/10.1007/JHEP08(2018)091}{\emph{JHEP} {\bf 08} (2018)
  091}, [\href{http://arxiv.org/abs/1802.03665}{{\tt 1802.03665}}].

\bibitem{Broedel:2009nsh}
J.~Broedel and L.~J. Dixon, \emph{{$R^4$ counterterm and $E_{7(7)}$ symmetry in
  maximal supergravity}},
  \href{http://dx.doi.org/10.1007/JHEP05(2010)003}{\emph{JHEP} {\bf 05} (2010)
  003}, [\href{http://arxiv.org/abs/0911.5704}{{\tt 0911.5704}}].

\bibitem{Elvang:2010jv}
H.~Elvang, D.~Z. Freedman and M.~Kiermaier, \emph{{A simple approach to
  counterterms in N=8 supergravity}},
  \href{http://dx.doi.org/10.1007/JHEP11(2010)016}{\emph{JHEP} {\bf 11} (2010)
  016}, [\href{http://arxiv.org/abs/1003.5018}{{\tt 1003.5018}}].

\bibitem{Beisert:2010jx}
N.~Beisert, H.~Elvang, D.~Z. Freedman, M.~Kiermaier, A.~Morales and
  S.~Stieberger, \emph{{$E_{7(7)}$ constraints on counterterms in N=8
  supergravity}},
  \href{http://dx.doi.org/10.1016/j.physletb.2010.09.069}{\emph{Phys. Lett.}
  {\bf B694} (2011) 265--271}, [\href{http://arxiv.org/abs/1009.1643}{{\tt
  1009.1643}}].

\bibitem{Freedman:2011uc}
D.~Z. Freedman and E.~Tonni, \emph{{The $D^{2k} R^4$ Invariants of N=8
  Supergravity}}, \href{http://dx.doi.org/10.1007/JHEP04(2011)006}{\emph{JHEP}
  {\bf 04} (2011) 006}, [\href{http://arxiv.org/abs/1101.1672}{{\tt
  1101.1672}}].

\bibitem{Freedman:2018mrv}
D.~Z. Freedman, R.~Kallosh and Y.~Yamada, \emph{{Duality Constraints on
  Counterterms in $\mathcal N=5,\ 6$ Supergravities}},
  \href{http://dx.doi.org/10.1002/prop.201800054}{\emph{Fortsch. Phys.} {\bf
  2018} (2018) 1800054}, [\href{http://arxiv.org/abs/1807.06704}{{\tt
  1807.06704}}].

\bibitem{Bossard:2009sy}
G.~Bossard, P.~S. Howe and K.~S. Stelle, \emph{{The Ultra-violet question in
  maximally supersymmetric field theories}},
  \href{http://dx.doi.org/10.1007/s10714-009-0775-0}{\emph{Gen. Rel. Grav.}
  {\bf 41} (2009) 919--981}, [\href{http://arxiv.org/abs/0901.4661}{{\tt
  0901.4661}}].

\bibitem{Bern:1998ug}
Z.~Bern, L.~J. Dixon, D.~C. Dunbar, M.~Perelstein and J.~S. Rozowsky, \emph{{On
  the relationship between Yang-Mills theory and gravity and its implication
  for ultraviolet divergences}},
  \href{http://dx.doi.org/10.1016/S0550-3213(98)00420-9}{\emph{Nucl. Phys. B}
  {\bf 530} (1998) 401--456}, [\href{http://arxiv.org/abs/hep-th/9802162}{{\tt
  hep-th/9802162}}].

\bibitem{Bern:2012cd}
Z.~Bern, S.~Davies, T.~Dennen and Y.-t. Huang, \emph{{Absence of Three-Loop
  Four-Point Divergences in N=4 Supergravity}},
  \href{http://dx.doi.org/10.1103/PhysRevLett.108.201301}{\emph{Phys. Rev.
  Lett.} {\bf 108} (2012) 201301}, [\href{http://arxiv.org/abs/1202.3423}{{\tt
  1202.3423}}].

\bibitem{Carrasco:2013ypa}
J.~J.~M. Carrasco, R.~Kallosh, R.~Roiban and A.~A. Tseytlin, \emph{{On the U(1)
  duality anomaly and the S-matrix of N=4 supergravity}},
  \href{http://dx.doi.org/10.1007/JHEP07(2013)029}{\emph{JHEP} {\bf 07} (2013)
  029}, [\href{http://arxiv.org/abs/1303.6219}{{\tt 1303.6219}}].

\bibitem{Bern:2013uka}
Z.~Bern, S.~Davies, T.~Dennen, A.~V. Smirnov and V.~A. Smirnov,
  \emph{{Ultraviolet Properties of N=4 Supergravity at Four Loops}},
  \href{http://dx.doi.org/10.1103/PhysRevLett.111.231302}{\emph{Phys. Rev.
  Lett.} {\bf 111} (2013) 231302}, [\href{http://arxiv.org/abs/1309.2498}{{\tt
  1309.2498}}].

\end{thebibliography}\endgroup

\end{document}